\documentclass{article}[12pt]
\usepackage{amsmath,amssymb}
\usepackage{graphicx}%
\usepackage{xcolor}
\usepackage{verbatim}
\usepackage{multirow}

\newtheorem{stat}{Statement}

\newcommand{\dt}[1]{\frac{d#1}{dt}}

\newcommand{\Fig}[4]{%
\begin{center}
\parbox{#2cm}{%
\includegraphics[width=#2cm,height=#3cm]{#1}}\\
\parbox{14cm}{\noindent\refstepcounter{figure} Figure: \thefigure:\quad #4}\end{center}}

\newcommand{\TwoFigs}[4]{%
\begin{flushleft}
\begin{tabular}{cc}
\parbox{7cm}{\centerline{\includegraphics[width=7cm]{#1}}}  & \parbox{7cm}{\centerline{\includegraphics[width=7cm]{#2}}}  \\
\parbox{7cm}{\vspace{7pt}\refstepcounter{figure} Figure: \thefigure.\quad #3\vfill} & \parbox{7cm}{\vspace{7pt}\refstepcounter{figure} Figure: \thefigure.\quad #4\vfill}\\
\end{tabular}
\end{flushleft}
\vspace{7pt}
}
\newcommand{\TwoFigsReg}[6]{%
\begin{flushleft}
\begin{tabular}{cc}
\parbox{7.5cm}{\centerline{\includegraphics[width=7cm,height=#2cm]{#1}}}  & \parbox{7.5cm}{\centerline{\includegraphics[width=7cm,height=#4cm]{#3}}}  \\
\parbox{7.5cm}{\vspace{7pt}\refstepcounter{figure} Figure: \thefigure.\quad #5\vfill} & \parbox{7.5cm}{\vspace{7pt}\refstepcounter{figure} Figure: \thefigure.\quad #6\vfill}\\
\end{tabular}
\end{flushleft}
\vspace{7pt}
}

\newcounter{strochka}

\newcounter{spisok}
\begin{document}

\begin{center}
{\bf \Large Yu.G. Ignat'ev}\\
\footnote{Institute of Physics, Kazan Federal University, Kremlyovskaya str., 35, Kazan, 420008, Russia}\\[12pt]
{\bf \Large Evolution of plane perturbations in the cosmological environment of the Higgs scalar field and an ideal scalar charged fluid\\}
\end{center}

\abstract{A phenomenological model of an ideal fluid with a scalar charge is formulated, on the basis of which a model with a neutral fluid and a vacuum-field model with rules of transition between them are constructed. A qualitative analysis of the obtained dynamic systems is carried out and numerical cosmological models based on these systems are constructed. A mathematical model of plane longitudinal scalar-gravitational perturbations of the Friedmann ideal charged fluid with Higgs interaction is formulated. It is shown that in the absence of fluid, i.e., in the vacuum-field model, gravitational perturbations do not arise. Perturbations of the scalar field are possible only in those cases when in the unperturbed state the cosmological system is at singular points. For these cases, exact solutions of the field equation are found, expressed in Bessel functions of the first and second kind and describing damped oscillations in the case of a stable unperturbed state and growing oscillations in the case of an unstable unperturbed state. The WKB theory of plane scalar-gravitational perturbations is constructed: dispersion equations are obtained in general form and solved for a neutral fluid. In this case, expressions are obtained for the local frequency and growth increment of oscillations, as well as the integral increment. It is shown that only free wave regimes or growing standing oscillations are possible during the evolution. Perturbations in the WKB approximation in a neutral fluid are studied and it is shown that local formulas for the evolution of perturbations correspond to the model of the 1985 article by M.Yu. The times of the beginning and end of the instability phase are determined and it is shown that instability can develop only at the unstable inflationary stage of the expansion of the Universe. \\

}


\section{Introduction}
At present, the existence of supermassive black holes with masses of $10^9\div10^{11}M_\odot$ in the centers of galaxies and quasars has been reliably established (\cite{SMBH1}, \cite{SMBH2}). Supermassive black holes with masses of the order of $\sim10^9M_\odot$ are the central objects of luminous quasars observed at redshifts $z>6$. More than 200 quasars with $z> 6$ and several objects with $z> 7$ have been discovered to date. The quasar with the highest redshift at $z = 7.5$ has an absolute luminosity of $1.4\cdot10^{47}$ erg/sec and a mass of the central black hole of $1.6\pm0.4\cdot10^9 M_\odot$ \cite{Fan}. The astrophysical origin of such supermassive black holes in the early Universe remains poorly understood, as the standard gas accretion theory used in the theory of star and galaxy formation is unable to explain the very rapid growth of such objects in the early Universe.

The results of numerical simulations \cite{Trakhtenbrot} impose a number of constraints on the formation parameters of supermassive black holes. Light black hole embryos with a mass of $M\leqslant 10^3 M_\odot$ cannot grow to masses of the order of $10^8 M_\odot$ at $z = 6$ even under supercritical accretion. To form supermassive black holes with masses of $10^8 \div 10^9 M_\odot$, heavier embryos of $M\sim 10^4 \div 10^6 M_\odot$ and gas-rich galaxies containing quasars are necessary. However, at present there are no convincing models for the appearance of such heavy embryos in the early Universe. Furthermore, it was found that the spatial density of luminous quasars decreases rapidly with increasing redshift, with this trend becoming stronger beyond $z = 5\div 6$ \cite{Zhu}.

Interest in the mechanisms of formation of supermassive Black Holes, taking into account the fact of their dominant presence in quasars, is caused, in particular, by the fact that such Black Holes are formed in quasars at fairly early stages of the evolution of the Universe, before the formation of stars. This circumstance, in particular, opens up the possibility of formation of supermassive Black Holes under conditions when scalar fields and baryonic dark matter can have a significant impact on this process. In this regard, we note the works \cite{Supermass_BH} -- \cite{Soliton}, which consider the possibility of the existence of \emph{scalar halos} and \emph{scalar hair} in the vicinity of supermassive Black Holes.

Apparently, the scalar field plays a key role in the mechanism of formation of supermassive Black Holes in the early Universe. In the work \cite{Ign_GC21_Un}, based on the complete theory of a two-component system of degenerate scalar charged fermions with Higgs scalar fields \cite{TMF_21} and the results of numerical simulation of the corresponding cosmological model, a numerical model of the evolution of scalar-gravitational perturbations for the case of an asymmetric scalar doublet was constructed, examples of instability development in a cosmological system were given, and some features of this process were revealed. Further, in the works \cite{Yu_GC_3_22} -- \cite{YU_GC_4_22} a systematic study of the development of scalar-gravitational perturbations in a cosmological model based on a one-component system of degenerate scalar charged fermions with classical Higgs interaction was carried out for the possibility of the formation of supermassive black holes in the early Universe. In this case, the processes of their Houging evaporation were taken into account. The studies confirmed the fundamental possibility of the early formation of black holes with the necessary masses.

Further, in order to clarify the connection between instability and bifurcation points of the vacuum-scalar cosmological model, the influence of the phantom field on the process of gravitational instability development was investigated in \cite{Yu_GC_23}. It was shown that scalar-gravitational instability at early stages of expansion in the model under study occurs at sufficiently large scalar charges, and the instability develops precisely near unstable points of the vacuum doublet. In this case, short-wave perturbations of even a free phantom field turn out to be stable at stable singular points of the vacuum doublet. The key point of the theory of scalar-gravitational instability proposed in \cite{Yu_GC_3_22} -- \cite{Yu_GC_23} is the exponentially rapid growth of linear perturbations with time, which, in contrast to the standard Lifshitz power law for gas-liquid models, allows perturbations to grow to fairly large values ??over a very short interval of cosmological time. The reason for the rapid development of perturbations is apparently a combination of two factors: the interparticle scalar attraction of charged fermions upon reaching critical densities and the macroscopic gravitational attraction.

Although the above works substantiate the fundamental possibility of early formation of supermassive black holes, the corresponding hypothesis requires additional substantiation, in particular, the study of the evolution of spherical perturbations, which should correspond to the processes of formation of black holes. This problem was solved in the article \cite{TMF_23_1}. Note that an important role in the dissemination of the results of numerical simulation from a certain basic model to models with real physical parameters is played by the scale invariance of the theory with respect to similarity transformations of its fundamental parameters \cite{Trans}.

Note, however, that the theory of scalar-gravitational instability of the statistical system of scalar charged fermions constructed by the Author has a significant drawback - the extreme cumbersomeness of the coefficients of the perturbation evolution equations, which extremely complicates the analytical study of the model and leads to the need for its numerical integration. This factor deprives the theoretical model of physical transparency and prevents the prediction of its properties, which undoubtedly reduces its value. The aim of this work is, firstly, to construct a phenomenological model of a cosmological medium that adequately reflects the properties of the \emph{microscopically substantiated} model of scalar charged fermionic matter, and to establish the rules for the transition between different types of models. The second aim of the work is to construct a theory of flat gravitational-scalar perturbations of the isotropic Friedmann Universe determined by this medium and to study the limiting cases of this theory. The third aim of the work is to construct methods for WKB analysis of flat perturbations.

It should be noted that the motivating reasons for this study, among others, were, firstly, a discussion with Academician of the Russian Academy of Sciences \fbox{A. A. Starobinsky} in November 2023, as well as a reference in December 2023 by Professor D. Yu. Khlopov to the 1985 work \cite{Khlopov}, which investigated the instability of scalar perturbations in connection with the problem of the formation of primordial Black Holes. The author did not even suspect the existence of this interesting work due to the long time since its publication.

\section{Mathematical model of the cosmological system \newline of an ideal fluid with a Higgs scalar field}
\subsection{Self-consistent system of equations}
Unlike \cite{TMF_21}, in which a macroscopic model of a statistical system of scalar charged particles described by macroscopic flows can be formulated based on the Lagrangian formalism from the microscopic equations of motion of scalar charged particles, in this article we will consider a simpler phenomenological model of matter based on the classical Higgs field $\Phi$ and the energy-momentum tensor of an ideal fluid.
The Lagrange function $L_s$ of the scalar Higgs field is \footnote{Here and below, Latin letters run through the values $\overline{1,4}$, Greek letters -- $\overline{1,3}$. The Planck system of units \newline $G=c=\hbar=1$ is used throughout. \label{Plank_units}}
\begin{eqnarray} \label{L_s}
L_s=\frac{1}{16\pi}(g^{ik} \Phi_{,i} \Phi_{,k} -2V(\Phi)),
\end{eqnarray}
where
\begin{eqnarray}
\label{Higgs}
V(\Phi)=-\frac{\alpha}{4} \left(\Phi^{2} -\frac{m^{2}}{\alpha}\right)^{2}
\end{eqnarray}
-- potential energy of a scalar field, $\alpha$ -- self-action constant, $m$ -- mass of quanta.
The energy-momentum tensor of scalar fields with respect to the Lagrange function \eqref{L_s} is:
\begin{eqnarray}\label{T_s}
S^i_{k} =\frac{1}{16\pi}\bigl(2\Phi^{,i}\Phi_{,k}- \delta^i_k\Phi_{,j} \Phi^{,j}+2V(\Phi)\delta^i_k \bigr),
\end{eqnarray}

By the standard procedure of variation of the Lagrange function $L_s$ \eqref{L_s} (see, for example, \cite{Land_Field}) we obtain the equation of the scalar field
\begin{equation}\label{Eq_S}
\triangle\Phi+V'_\Phi=-8\pi\sigma\Rightarrow \triangle\Phi+\Phi(m^2-\alpha\Phi^2)=-8\pi\sigma,
\end{equation}
where the scalar charge density $\sigma$ in the considered phenomenological model of matter, in contrast to the model based on the microscopic theory, remains an undefined scalar.
Further, the energy-momentum tensor of an ideal fluid is equal to:
\begin{equation}\label{T_p}
T^i_{k}=(\varepsilon+p)u^i u_k-\delta^i_k p,
\end{equation}
where $u^i$ is the unit time-like vector of the dynamic velocity of the fluid
\begin{equation}\label{u^2=1}
(u,u)=1.
\end{equation}

Einstein's equations for the system ``scalar field + ideal fluid'' have the form:
\begin{equation}\label{EQ_Einst}
R^i_k-\frac{1}{2}\delta^i_k R=8\pi (T^i_k+S^i_k) + \delta^i_k \Lambda_0,
\end{equation}
where $\Lambda_0$ is the seed value of the cosmological constant, related to its observed value $\Lambda$, obtained by removing the constant terms in the potential energy, by the relation:
\begin{equation}\label{lambda0->Lambda}
\Lambda=\Lambda_0-\frac{m^4}{4\alpha}.
\end{equation}

Calculating the covariant divergence of both parts of the Einstein equations \eqref{EQ_Einst}, we obtain the laws of conservation of energy - the momentum of the system:
\begin{equation}\label{2}
\nabla _{k} T^{ik}=\sigma\nabla^{i} \Phi =0.
\end{equation}

From the normalization relation of the velocity vector follows the well-known identity
\begin{equation}\label{6}
u^k_{~,i}u_k\equiv 0,
\end{equation}
which allows us to reduce the laws of conservation of energy - momentum (\ref{2}) to the form of equations of ideal hydrodynamics
\begin{eqnarray}\label{2a}
(\varepsilon+p)u^i_{~,k}u^k=(g^{ik}-u^iu^k)(p_{,k}+\sigma\Phi_{,k});\\
\label{2b}
\nabla_k[(\varepsilon+p)u^k]=u^k(p_{,k}+\sigma\Phi_{,k}).
\end{eqnarray}
\subsection{Equation of state}
In what follows, the scalar functions $\sigma$, $\varepsilon$ and $p$ included in the equation of the scalar field \eqref{Eq_S}, the definition of the energy-momentum tensor of an ideal fluid \eqref{T_p} and the equations of hydrodynamics \eqref{2a} -- \eqref{2b}, will be called \emph{macroscopic scalars}. In the phenomenological hydrodynamics of an ideal fluid, to obtain a closed self-consistent system of equations, equations of connection between similar scalars are introduced, which are called equations of state:
\begin{eqnarray}\label{p,s(e)}
p=p(\varepsilon),& \displaystyle \sigma=\sigma(\varepsilon).
\end{eqnarray}

In the case of a system of scalar charged particles, the coupling equations \eqref{p,s(e)} cannot be introduced. Indeed, the expressions for these macroscopic scalars, obtained on the basis of microscopic dynamics in our case of a scalar singlet, have the form \cite{TMF_21}:
\begin{eqnarray}
\label{6b}
\varepsilon=m_*^4\frac{2S+1}{2\pi^2}\int\limits_{0}^\infty \frac{\sqrt{1+z^2}z^2dz}{\exp(-\gamma+\lambda\sqrt{1+z^2})\pm1}\geqslant0;\\
\label{6c}
p=m_*^4\frac{2S+1}{6\pi^2}\int\limits_{0}^\infty \frac{1}{\sqrt{1+z^2}}\cdot\frac{z^4dz}{\exp(-\gamma+\lambda\sqrt{1+z^2})\pm1}\geqslant0;\\
\label{6e}
\sigma =m_*^3 q\frac{2S+1}{(2\pi)^{3}}
\int\limits_{0}^\infty\frac{1}{\sqrt{1+z^2}}\cdot\frac{z^2dz}{\exp(-\gamma+\lambda\sqrt{1+z^2})\pm1},
\end{eqnarray}
where $m_*=m_0+q\Phi$ is the dynamic mass of the particles ($m_0$ is the bare mass, $q$ is the scalar charge), $\lambda=|m_*|/\theta$, $\gamma=\mu/\theta$, ($\theta$ is the temperature. $\mu$ is the chemical potential), $S$ is the spin of the particles, and the signs of $\pm$ in the integrands correspond to fermions and bosons. Then from \eqref{6b} -- \eqref{6e} follows a strict relation between the scalar charge density $\sigma$ and the trace $T$ of the energy-momentum tensor of the liquid\footnote{This strict relation is already obtained at the level of microscopic particle dynamics before applying the operation of averaging over states \cite{TMF_21}.}:
\begin{equation}\label{sigma=}
\sigma=\frac{q}{m_*}T\equiv \frac{q}{m_*}(\varepsilon-3p)\Rightarrow p=\frac{1}{3}\biggl(\varepsilon-\frac{m_*}{q}\sigma\biggr),
\end{equation}
moreover
\begin{equation}\label{T>=0}
T\equiv T^i_i=\varepsilon-3p\geqslant0.
\end{equation}
In this case, generally speaking, there is no restriction on the sign of the scalar charge density $\sigma$. In the case of zero bare mass of scalar charges ($m_0=0$), we obtain from \eqref{sigma=} -- \eqref{T>=0}
\begin{eqnarray}\label{sigma1=}
\sigma=\frac{\varepsilon-3p}{\Phi}; & \displaystyle p=\frac{1}{3}(\varepsilon-\Phi\sigma)\Rightarrow \Phi\sigma\leqslant \varepsilon.
\end{eqnarray}

The relations \eqref{sigma=} and \eqref{sigma1=} demonstrate the impossibility of the relation of the type \eqref{p,s(e)} in the case of a system of scalar charged particles. In particular, the expression for the scalar charge density according to these relations is singular at $m_*=0$ or $\Phi=0$. This creates difficult-to-overcome obstacles to the study of a system of scalar charged particles with standard equations of state of the type \eqref{p,s(e)}. Thus, in the case of systems with scalar charged particles, we must modify the standard approach of ideal fluid hydrodynamics.

Note that for an equilibrium statistical system of fermions under conditions of complete degeneracy $\gamma\to\infty$, the macroscopic densities \eqref{6b} -- \eqref{6e} are expressed in elementary functions \cite{Ignat14_2}\footnote{In the case of $\gamma\to0$ and the Boltzmann distribution, these scalars are expressed in terms of modified Bessel functions.}:
\begin{eqnarray}\label{2_3}
\varepsilon = {\displaystyle \frac{m_*^4}{8\pi^2}}(2\psi^3\sqrt{1+\psi^2}+F_1(\psi)); \; 
p =\frac{m_*^4}{24\pi^2}(2\psi^3\sqrt{1+\psi^2}-3F_1(\psi)); \; \displaystyle (\varepsilon-3p)=\frac{m_*^4}{2\pi^2}F_1(\psi),
\end{eqnarray}
where the dimensionless function $ \psi=\pi_f/|m_*|$ is introduced, equal to the ratio of the Fermi momentum $\pi_f$ to the total energy of the fermion $|m_*|$, and its dimensionless function $F_1(\psi)$ (see \cite{Ignat14_2}):
\[ F_1(\psi)=\psi\sqrt{1+\psi^2}-\ln(\psi+\sqrt{1+\psi^2}).\]
From \eqref{2_3} follows the identity
\begin{equation}\label{E_P_f}
\varepsilon+p\equiv \frac{1}{3\pi^2}m_*^4\psi^3\sqrt{1+\psi^2}.\end{equation}

For $m\to 0\Leftrightarrow \psi\to\infty$, macroscopic scalars \eqref{2_3} have the following asymptotics:
\begin{eqnarray}\label{scal:m->0}
\left.\sigma\right|_{m_*\to0}\simeq\frac{qm_*\pi^2_f}{2\pi^2}\to 0; & \displaystyle \left.\varepsilon\right|_{m_*\to0}\simeq 3\left.p\right|_{m_*\to0}\simeq \frac{\pi^4_f}{4\pi^2}=\mathrm{Const}(m_*); & \displaystyle  \left.(\varepsilon-3p)\right|_{m_*\to0}=\frac{m_*^2\pi_f^2}{2\pi^2} ,
\end{eqnarray}
where $\mathrm{Const}(m_*)$ means the independence of the quantity from $m_*$.

Thus, the following statement is true:
\begin{stat}\label{sigma_not=8}
\hskip -6pt \textbf{.}\quad
When $m_*\to0$, the macroscopic scalars $n,p,\varepsilon$ are finite, and the scalar charge density $\sigma\sim m_*$ tends to zero, i.e.,
the macroscopic scalars are not singular when $m_*=0$.\\

Note that this property is universal and does not depend on the particle statistics or the chemical potential.
\end{stat}

In this connection, we will further introduce an equation of state that satisfies the above properties of macroscopic scalars:
\begin{equation}\label{sigma0=}
\sigma=\beta^2 m_* q\rho(\varepsilon);\quad m_0=0\Rightarrow \sigma=q^2\beta^2\Phi\rho(\varepsilon),
\end{equation}
where $\beta$ is some constant of dimension $1/q$ in the equation of state of the liquid, $\rho(\varepsilon)$ is some given non-negative function of the dimension of the energy density, so that
\begin{eqnarray}\label{p=p(rho)}
p=\frac{1}{3}\bigl(\varepsilon-\beta^2 m_*^2\rho(\varepsilon)\bigr);& \displaystyle m_0=0\Rightarrow p=\frac{1}{3}\bigl(\varepsilon-q^2\beta^2\Phi^2\rho(\varepsilon)\bigr), \\
\label{rho><}
0\leqslant \rho(\varepsilon)\leqslant \frac{\varepsilon}{\beta^2m_*^2}. &
\end{eqnarray}
Indeed, when $q\to0$
\begin{equation}\label{q->0}
q\to0\Rightarrow \sigma\to0,\quad p\to \frac{1}{3}\varepsilon
\end{equation}
-- in full accordance with the asymptotic properties of \eqref{scal:m->0}, i.e., $q\to0$ corresponds to the ultrarelativistic equation of state of the liquid, as it should be \cite{TMF_21}.

Given a linear relation
\[\rho(\varepsilon) =\varkappa\varepsilon,\]
comparing, for example, \eqref{sigma0=} with \eqref{scal:m->0}, we find an expression for $\beta^2$ in the case of a completely degenerate system of scalar charged fermions:
\begin{equation}\label{beta=}
\varkappa\beta^2=\frac{2}{\pi^2_f}\Rightarrow \varkappa e^2=2\left(\frac{q}{\pi_f}\right)^2.
\end{equation}

\noindent Thus, the following statement is true:
\begin{stat}\label{div}
\hskip -6pt \textbf{.}\quad The fundamental difference between the phenomenological hydrodynamics of an ideal scalar charged liquid and the phenomenological hydrodynamics of an ideal neutral liquid is the establishment of an equation of state between the scalar charge density and the energy density, and not between the pressure and the energy density.

The transition to a system of scalar neutral particles is accomplished, firstly, by substituting $q=0$, and secondly, by using the equation of state $p=p(\varepsilon)$ instead of the equation of state \eqref{sigma0=}.
\end{stat}

\section{Background solution\label{Background}}
\subsection{Background dynamic system}
As a background, we consider the spatially flat Friedman metric
\begin{eqnarray}\label{ds_0}
ds_0^2=dt^2-a^2(t)(dx^2+dy^2+dz^2)\equiv 
dt^2-a^2(t)[dr^2+r^2(d\theta^2+\sin^2\theta d\varphi^2)],
\end{eqnarray}
and as a background solution we consider a homogeneous isotropic distribution of matter, in which all hydrodynamic functions and the scalar field depend only on the cosmological time $t$:
\begin{equation}\label{base_state}
\Phi=\Phi(t);\;  \varepsilon=\varepsilon(t);\; p=p(t);\;  u^i=u^i(t).
\end{equation}
Note that the physically measurable radius in the \eqref{ds_0} metric is
\begin{equation}\label{R=}
R=a(t)r.
\end{equation}
It is easy to verify that
\begin{equation}\label{u_0}
u^i=\delta^i_4
\end{equation}
turns equations (\ref{2a}) into identities, and equation (\ref{2b}) reduces to the \emph{material} equation\footnote{Here and below $\dot{f}=\partial f/\partial t$, $f'=\partial f/\partial t$.}
\begin{equation}\label{7a1-0}
\dt{\varepsilon}+3\frac{\dot{a}}{a}(\varepsilon+p)=\sigma\dot{\Phi};
\end{equation}

Further, the energy-momentum tensor of the scalar field in the background state also takes the form of the energy-momentum tensor of an ideal isotropic liquid:
\begin{equation} \label{MET_s}
S^{ik} =(\varepsilon_s +p_{s} )u^{i} u^{k} -p_s g^{ik} , \end{equation} 
where: 
\begin{eqnarray}\label{Es} \varepsilon_s=\frac{1}{8\pi}\biggl(\frac{1}{2}\dot{\Phi}^2+V(\Phi)\biggr);\\ \label{Ps} p_{s}=\frac{1}{8\pi}\biggl(\frac{1}{2} \dot{\Phi}^2-V(\Phi)\biggr), \end{eqnarray} 
so: 
\begin{equation}\label{e+p} \varepsilon_s+p_{s}=\frac{1}{8\pi}\dot{\Phi}^2.
\end{equation}
The equations of the background scalar field \eqref{Eq_S} in the Friedmann metric take the form:
\begin{eqnarray}\label{Eq_Phi_eta}
\ddot{\Phi}+\frac{3}{a}\dot{a}\dot{\Phi}+m_0^2\Phi-\alpha\Phi^3= -8\pi\sigma.
\end{eqnarray}
Assuming further that the equation of state $\rho=\rho(\varepsilon)$ \eqref{sigma=} and the expression for pressure \eqref{p=p(rho)} are valid, taking into account the definition of the material equation \eqref{7a1-0} we obtain the complete system of dynamic equations of the background model:
\begin{eqnarray}\label{Dxi/dt}
\dot{\xi}=H;\\
\label{dH/dt}
\dot{H}=- \frac{Z^2}{2}-\frac{4\pi}{3}\bigl(4\varepsilon-\beta^2q^2\Phi^2\rho(\varepsilon)\bigr);\\
\label{dPhi/dt}
\dot{\Phi}=Z;\;\\
\label{dZ/dt}
\dot{Z}=-3HZ-\Phi(m^2 -\alpha\Phi^2)-8\pi \beta^2q^2\Phi\rho(\varepsilon);\\
\label{de/dt}
\dot{\varepsilon}=\displaystyle -H\bigl(4\varepsilon-\beta^2q^2\Phi^2\rho(\varepsilon)\bigr)+e^2\Phi Z\rho(\varepsilon),
\end{eqnarray}
where the Hubble parameter $H(t)$ is introduced
\begin{equation}\label{H}
H= \frac{\dot{a}}{a}\equiv \dot{\xi}.
\end{equation}
In this case, Einstein's equation $^4_4$, which is the first integral of the system \eqref{Dxi/dt} -- \eqref{de/dt} with a zero value of the constant, takes the form
\begin{eqnarray}\label{Surf_Einst}
3H^2-\frac{Z^2}{2}-8\pi\varepsilon+\frac{\alpha}{4}\left(\Phi^2-\frac{m^2}{\alpha}\right)^2-\Lambda_0=0.
\end{eqnarray}

Note that the cosmological constant $\Lambda_0$ is present only in the equation \eqref{Surf_Einst}, thereby defining the first integral of the dynamic system \eqref{Dxi/dt} -- \eqref{de/dt}.
For a given equation of state $\rho=\rho(\varepsilon)$, the system of background equations \eqref{Dxi/dt} -- \eqref{de/dt} is a completely defined autonomous normal system of differential equations $\mathbf{S_5}$, representing the dynamic system in the five-dimensional arithmetic phase space $\mathbb{R}_5=\{\xi,H, \Phi,Z,\varepsilon\}$. Its subsystem $\mathbf{S_4}$: \eqref{dH/dt} -- \eqref{de/dt}, obtained by removing the equation \eqref{Dxi/dt}, is also an autonomous dynamic system in the four-dimensional phase space $\mathbb{R}_4=\{H, \Phi,Z,\varepsilon\}$. This system is convenient for qualitative analysis. The phase trajectories of the system
$\mathbf{S_4}$ lie on the Einstein-Higgs hypersurface \eqref{Surf_Einst}.

The corresponding system of dynamic equations in the case of a scalar neutral fluid, taking into account the Statement \ref{div}, takes the form:
\begin{eqnarray}\label{Dxi/dt00}
\dot{\xi}=H;\\
\label{dH/dt00}
\dot{H}=- \frac{Z^2}{2}-4\pi(\varepsilon+p(\varepsilon));\\
\label{dPhi/dt00}
\dot{\Phi}=Z;\;\\
\label{dZ/dt00}
\dot{Z}=-3HZ-\Phi(m^2 -\alpha\Phi^2)=0;\\
\label{de/dt00}
\dot{\varepsilon}=\displaystyle -(\varepsilon+p(\varepsilon))H=0.
\end{eqnarray}
\subsection{Numerical modeling of the background state}
\subsubsection{Background model with a scalar charged ideal fluid: $\mathbf{S_5}$}
For numerical modeling of the background state, we consider the barotropic equation of state between the scalar charge density and the energy density of the fluid with \emph{barotropic coefficient} $\varkappa$:
\begin{equation}\label{p=ke}
\rho=\varkappa\varepsilon \qquad (\varkappa \geqslant0).
\end{equation}
Thus, we obtain from the system of equations
\begin{eqnarray}\label{Dxi/dt0}
\dot{\xi}=H;\\
\label{dPhi/dt0}
\dot{\Phi}=Z;\;\\
\label{dH/dt_0}
\dot{H}=- \frac{Z^2}{2}-\frac{4\pi}{3}\varepsilon(4-e^2\Phi^2);\\
\label{dZ/dt0}
\dot{Z}=-3HZ-\Phi(m^2 -\alpha\Phi^2)-8\pi e^2\Phi\varepsilon;\\
\label{de/dt0}
\dot{\varepsilon}=\displaystyle -\varepsilon(4-e^2\Phi)H+e^2\Phi Z\varepsilon,
\end{eqnarray}
where a new dimensionless parameter $e$ is introduced -- \emph{dimensionless scalar charge}
\begin{equation}\label{e=}
e^2\equiv\varkappa \beta^2 q^2\geqslant0,
\end{equation}
through which, according to \eqref{sigma0=} and \eqref{p=p(rho)}, the scalar charge density and liquid pressure are expressed:
\begin{eqnarray}\label{p,sigma=}
\sigma=e^2\Phi\varepsilon, & \displaystyle p=\frac{1}{3}(1-e^2\Phi^2)\varepsilon.
\end{eqnarray}

Using the autonomy of this system and the admissible transformations of the Friedman metric, one can always choose the initial conditions in the form:
\begin{equation}\label{xi(0)=0}
\xi(0)=0\Rightarrow a(0)=1.
\end{equation}
Let further $\Phi_0,Z_0,\varepsilon_0$ be the initial values ??of the corresponding quantities. Then the initial value $H(0)$ is defined as one of the symmetric roots $H_\pm$ of the equation of the Einstein-Higgs surface \eqref{Surf_Einst}:
\begin{eqnarray}
H_\pm\equiv \pm H_0 =\pm\frac{1}{\sqrt{3}}\cdot\sqrt{\frac{Z_0^2}{2}-\frac{\alpha}{4}\left(\Phi_0^2-\frac{m^2}{\alpha}\right)^2+8\pi\varepsilon_0+\Lambda_0}.
\end{eqnarray}
From this it is clear that not all initial conditions on the functions $\Phi,Z,\varepsilon$ can be used in the Cauchy problem, but only those that satisfy the \emph{energy condition} $H^2\geqslant0$, i.e., only those that lie outside the energy surface $H^2=0$, representing a deformed hyperbolic cone
\begin{equation}\label{H2>0$}
\frac{Z_0^2}{2}-\frac{\alpha}{4}\left(\Phi_0^2-\frac{m^2}{\alpha}\right)^2+8\pi\varepsilon_0+\Lambda_0\geqslant0.
\end{equation}
Taking this condition into account, we define the initial conditions of the Cauchy problem for the system \eqref{Dxi/dt0} -- \eqref{de/dt} in the form of an ordered list:
\begin{equation}\label{IC}
[\xi(0)=0,H(0)=\epsilon H_0,\Phi(0)=\Phi_0,Z(0)=Z_0,\varepsilon(0)=\varepsilon_0] \Rightarrow \textbf{I}=[\Phi_0,Z_0,\varepsilon_0,\epsilon],
\end{equation}
where $\epsilon=\pm1$ -- the Cauchy problem is completely determined by the initial values ??of the three functions and the sign of the initial value of the Hubble parameter. In addition, the solution is determined by the values ??of the four fundamental parameters of the model
\begin{equation}\label{Param}
\mathbf{P}=[\alpha,m,e,\Lambda_0].
\end{equation}
By removing the equation \eqref{Dxi/dt0} from the dynamic system $\mathbf{S_5}$, we obtain the dynamic system $\mathbf{S_4}$.

\subsubsection{Background model with neutral ideal fluid: $\mathbf{S^{(0)}_5}$}
Along with this model, we will consider a similar model with a neutral fluid with a barotropic coefficient $k$ ($p=k\varepsilon$), $\mathbf{S^{(0)}_5}$:
\begin{eqnarray}\label{Dxi/dt00_k}
\dot{\xi}=H;\\
\label{dH/dt00_k}
\dot{H}=- \frac{Z^2}{2}-4\pi\varepsilon(1+k);\\
\label{dPhi/dt00_k}
\dot{\Phi}=Z;\;\\
\label{dZ/dt00_k}
\dot{Z}=-3HZ-\Phi(m^2 -\alpha\Phi^2)=0;\\
\label{de/dt00_k}
\dot{\varepsilon}=\displaystyle -\varepsilon(1+k))H=0\Rightarrow \varepsilon=\varepsilon_0 a^{-(1+k)}.
\end{eqnarray}
The equation of the Einstein-Higgs hypersurface \eqref{Surf_Einst} in this model does not formally change.

\subsubsection{Vacuum-field background model: $\mathbf{S^{(00)}_4}$}
The vacuum-field background model $\mathbf{S^{(00)}_4}$ is obtained from the model $\mathbf{S_5}$ subject to the condition
\begin{equation}\label{vac_mod}
\mathbf{S^{(00)}_4}: \qquad \varepsilon\equiv 0:
\end{equation}
\begin{eqnarray}\label{Dxi/dt00}
\dot{\xi}=H; \quad \dot{H}=- \frac{Z^2}{2};\quad \dot{\Phi}=Z;\quad \dot{Z}=-3HZ-\Phi(m^2 -\alpha\Phi^2)=0.
\end{eqnarray}
By removing the equation on $\xi$ we obtain a three-dimensional autonomous dynamic system $\mathbf{S^{(00)}_3}$ with a two-dimensional Einstein-Higgs hypersurface.

\subsubsection{Dimensional analysis of the model}
Let us analyze the dimensions of the quantities that define the mathematical model. From the definitions \eqref{6b} -- \eqref{6e} it follows that the macroscopic scalars $\varepsilon$, $p$ and $\sigma$ have the same dimension, and from the Einstein equations \eqref{EQ_Einst} it follows that this dimension is equal to $1/\ell^2$, where $\ell$ is the dimension of the length
\[ [\varepsilon]=[p]=[\sigma]=[\Lambda]=[\ell^{-2}].\]
Further, from \eqref{T_s} and Einstein's equations \eqref{EQ_Einst} it follows
\[ [\Phi]=[\xi]=[1];\; [m]=[Z]=[H]=[\ell^{-1}];\; [\alpha]=[\ell^{-2}];\; [q]=[\pi_f]=[\ell^{-1/2}].\]
But then from the definitions of the function $\rho(\varepsilon)$ \eqref{sigma0=} and \eqref{e=} it follows
\[ [\beta]=[\ell^{1/2}];\; [e]=[1];\;  [\rho]=[\ell^{-2}];\; [\varkappa]=[1].\]
\subsubsection{Similarity transformations of a cosmological model}
By analogy with \cite{Trans} and the above dimensional analysis, we can formulate and prove the transformation properties of the model under study:

\begin{stat}
Let us consider two cosmological models: $\mathbf{M}$ with fundamental parameters $\mathbf{P}$ and initial conditions $\mathbf{I}$
and a similar model $\tilde{\mathbf{M}}$ \emph{with similarity coefficient} $\zeta$ with fundamental parameters $\tilde{\mathbf{P}}$ and initial conditions $\tilde{\mathbf{I}}$ -- 
\begin{eqnarray}\label{Par_tilde} \tilde{\mathbf{P}}=\biggl[\frac{\alpha}{\zeta^2},\frac{m}{\zeta},e,\frac{\Lambda}{\zeta^2}\biggr];\\ \label{Inits_tilde} \tilde{\mathbf{I}}=\biggl[\ Phi_0,\frac{1}{\zeta}Z_0,\frac{1}{\zeta^2}\varepsilon_0,\epsilon \biggl] .
\end{eqnarray} 
\begin{equation}\label{tilde_t} \tilde{t}=\zeta t.
\end{equation}
Let the solutions of the dynamic system $\mathbf{S_5}$ \eqref{Dxi/dt} -- \eqref{de/dt} for the model $\mathbf{M}$ \eqref{IC} -- \eqref{Param} be
\[\mathbf{S}(t)=[\xi(t),H(t),\Phi(t),Z(t),\varepsilon(t)].\]
Then the solutions of the corresponding equations for a similar model $\tilde{\mathbf{M}}$ \eqref{Inits_tilde} -- \eqref{Par_tilde} will be 
\begin{eqnarray}\label{Sol_tilde} \tilde{\mathbf{S}}(t)=[\tilde{\xi}(t),\tilde{H}(t),\tilde{\Phi}(t),\tilde{Z}(t),\tilde{\varepsilon}(t)]\equiv 
\biggl[\xi\biggl(\frac{t}{\zeta}\biggl),\frac{1}{\zeta}H\biggl(\frac{t}{k}\biggl),\Phi\biggl(\frac{t}{\zeta}\biggl),\frac{1}{\zeta}Z\biggl(\frac{t}{\zeta}\biggl),\frac{1}{\zeta^2}\varepsilon\biggl(\frac{t}{\zeta}\biggl)\biggr].
\end{eqnarray}
\end{stat}
This property of the system under study allows extending the results obtained with parameter values ??convenient for numerical modeling to models with real physical parameters, usually very small. The same applies to the dynamic system \eqref{Dxi/dt00} -- \eqref{de/dt00} with a neutral liquid, with the only difference being that $e$ and must be excluded from the formulas given above and the value $\varkappa$ in the parameters \eqref{Par_tilde} must be replaced by the barotropic coefficient $k$.

\subsection{Qualitative analysis of the background dynamic system}
The singular points of the background dynamic system $\mathbf{S_4}$ \eqref{dH/dt} -- \eqref{de/dt} are determined by the equality to zero of the right-hand sides of the corresponding dynamic equations (see, for example, \cite{Bogoyav}, \cite{Bautin}), which yields:
\begin{eqnarray}\label{sing_points}
1.\ Z=0;\quad 2.\ \varepsilon+p=0\Leftrightarrow 4\varepsilon-\beta^2q^2\Phi\rho(\varepsilon)=0;\quad 3.\ \Phi(m^2-\alpha\Phi^2)+8\pi \beta^2 q^2\Phi\rho=0.
\end{eqnarray}
From \eqref{sing_points}.2 and \eqref{T>=0} it follows that at singular points
\[\varepsilon=p=0\Rightarrow \rho=0.\]
but then from \eqref{sing_points}.3 it follows:
\[\Phi(m^2-\alpha\Phi^2)=0.\]

Thus, the studied dynamic system $\mathbf{S_4}$ \eqref{dH/dt} -- \eqref{de/dt} for $\Lambda>0\Rightarrow\Lambda_0>0$ can have 6 symmetric singular points:
\begin{eqnarray}\label{M_0}
M^\pm_0=\left[\pm \sqrt{\frac{\Lambda}{3}},0,0,0\right];&\displaystyle M^\pm_\pm=\left[\pm \sqrt{\frac{\Lambda_0}{3}},\pm\frac{m}{\sqrt{\alpha}},0,0\right],
\end{eqnarray}
where the signs before the radicals take on values ??independent of each other. The eigenvalues ??of the dynamic system matrix at these points are equal to:
\begin{eqnarray}
\label{lambda0}
M_0^\pm: & \displaystyle \lambda_0=& \displaystyle \biggl[-6H_0\equiv \mp6\sqrt{\frac{\Lambda}{3}},\quad -4H_0\equiv \mp4\sqrt{\frac{\Lambda}{3}}, \nonumber\\
 &  & -\frac{3}{2}H_0\pm \frac{1}{2}\sqrt{9H^2_0-4m^2}\quad \equiv \mp\frac{\sqrt{3\Lambda}}{2}\pm\frac{\sqrt{3\Lambda-4m^2}}{2}\ \biggr];\\
\label{lambda_pm}
M_\pm^\pm: & \displaystyle \lambda_\pm =& \displaystyle \biggl[-6H_\pm\equiv\mp6\sqrt{\frac{\Lambda_0}{3}}, \quad \left(4-\frac{e^2m^2}{\alpha}\right)\cdot H_\pm\equiv
\pm\left(4-\frac{e^2m^2}{\alpha}\right)\cdot\sqrt{\frac{\Lambda_0}{3}},\nonumber\\
 &  & \displaystyle -\frac{3}{2}H_\pm\pm\frac{1}{2}\sqrt{9H^2_\pm+8m^2} \quad \equiv \mp\frac{\sqrt{3\Lambda_0}}{2}\pm\frac{\sqrt{3\Lambda_0+8m^2}}{2}\ \biggr],
\end{eqnarray}
where again the signs before the radicals take on values ??independent of each other. The coordinates of the singular points coincide with the corresponding values ??for the \emph{vacuum-field model} $\mathbf{S^{(00)}_4}$ \cite{GC_20}, in which there is no ideal fluid.

Further, for $\Lambda\geqslant0\Rightarrow\Lambda_0\geqslant0$ the eigenvalues ??$\lambda_\pm$ are real and have two pairs of equal and opposite values, which correspond to saddle and nodal points. For $3\Lambda-4m^2>0$ the points $M^\pm_0$ have a similar character, and for $3\Lambda-4m^2<0$ the points $M^\pm_0$ become repulsive and attractive foci. Thus, at
\begin{equation}\label{attract_eq}
3\Lambda^2-4m^2<0,\qquad H_0=+\sqrt{\frac{\Lambda}{3}}
\end{equation}
there can be only one attracting stable singular point $M^+_0$.

The singular points of the system $\mathbf{S^{(0)}_4}$ \eqref{dH/dt00} -- \eqref{de/dt00} coincide with the singular points of the background dynamic system $\mathbf{S_4}$ \eqref{dH/dt} -- \eqref{de/dt}. The eigenvalues ??at these points are:
\begin{eqnarray}
\label{lambda00}
M_0^\pm: & \displaystyle \lambda_0=&\mp6\sqrt{\frac{\Lambda}{3}};\quad \mp (1+k)\sqrt{\frac{\Lambda}{3}}; \quad \mp\frac{\sqrt{3\Lambda}}{2}\pm\frac{\sqrt{3\Lambda-4m^2}}{2};\\ \label{lambda_pm0} M_\pm^\pm: & \displaystyle \lambda_\pm=&\mp6\sqrt{\frac{\Lambda_0}{3}};\quad \mp (1+k) \sqrt{\frac{\Lambda_0}{3}}; \quad\mp\frac{\sqrt{3\Lambda_0}}{2}\pm\frac{\sqrt{3\Lambda_0+8m^2}}{2}.
\end{eqnarray}
Note that the dynamic systems $\mathbf{S_4}$ and $\mathbf{S^{(0)}_4}$ differ only in the eigenvalues ??associated with the equation of state. Note also that the signs in the above formulas are associated, firstly, with the signs of the coordinates of the singular points, and, secondly, with the signs of the roots of the characteristic equation. In this case, each singular point corresponds to 4 eigenvalues.

\section{Numerical simulation of the background\label{Mod_Background}}
\subsection{Initial conditions at singular points}
Consider the following parameters for cosmological models with neutral fluid $\mathbf{S^{(0)}}$ and with charged fluid $\mathbf{S}$
\begin{eqnarray}\label{P_00}
\mathbf{P^{(0)}_0}=[1,1,1/3,1], & \mathbf{P_0}=[1,1,10^{-5},1].
\end{eqnarray}
According to \eqref{Param} and \eqref{Param}, we have the following parameter values ??for the models under study: $\alpha=1$, $m=1$, $e=10^{-5}$, $k=1/3$, $\Lambda_0=1\Rightarrow \Lambda=3/4$. When transforming the similarity \eqref{Par_tilde} -- \eqref{tilde_t} to realistic parameters of the SU(5) model with the similarity coefficient $\zeta=10^5$, we obtain from here:
\begin{equation}\label{SU5}
\tilde{\alpha}=10^{-10},\;\tilde{m}=10^{-5},\;\tilde{e}=10^{-5},\; \tilde{\Lambda}_0=10^{-10};\; \tilde{t}=10^5 t.
\end{equation}
Note, firstly, that the order of the parameter $e$ is chosen according to the results of \cite{Yu_GC_3_22}, \cite{TMF_23_1} -- this parameter should be small enough, secondly, according to \eqref{p=p(rho)} and \eqref{q->0} for such small values ??of the scalar charge $p\to1/3$ -- the equation of state of a scalar charged liquid tends to be ultrarelativistic, therefore in this sense the models $\mathbf{S^{(0)}_0}$ and $\mathbf{S_0}$ are close.

According to \eqref{M_0}, \eqref{lambda0} and \eqref{lambda_pm} the coordinates of the singular points and the corresponding eigenvalues ??in this case are:
\begin{eqnarray}\label{M^pm_00}
M^\pm_0=[\pm0.5,0,0,0]: &\displaystyle \lambda_0=[\mp3,\mp2,\mp0.75\pm i\cdot 0.661];\\
\label{M^pm_pm0}
M^\pm_\pm=[\pm0.577,\pm1,0,0]; & \displaystyle \lambda_\pm=[\mp 3.464,\mp 2.309, \mp 1.411, \pm 0.0886].
\end{eqnarray}
We first examine the case when the cosmological model is initially located at the saddle singular point $M^+_+$ ($\Phi_0=+1,\ \varepsilon_0=0.2$):
\begin{equation}\label{I_0}
\mathbf{I_0}=[1,0,0.2,1],\qquad  (H_0\approx 0.577).
\end{equation}

Fig. \ref{fig1} -- \ref{fig2} shows the evolution of the scale functions $\xi(t)$ and $H(t)$ of the $\mathbf{S^{(0)}}$ and $\mathbf{S}$ models for the case under study.
Both models have an initial singularity at time $t_0\approx-0.37449328$ and an infinite inflationary future. After exiting the singularity, the models quickly enter the inflationary expansion stage $H=0.577$, corresponding to the unstable saddle point $M^+_+$. In this case, the $\mathbf{S}$ model after a short stage of the first inflation passes into a stable attractive point with a zero scalar potential and the Hubble parameter $H=0.5$, while the model with a neutral liquid $\mathbf{S^{(0)}}$, unlike the $\mathbf{S_0}$ model, remains at the stage of the first inflation. This is apparently due to the fact that the subsystem \eqref{dPhi/dt00_k}--\eqref{dZ/dt00_k} in this case does not depend on $\varepsilon$ and for any $H(t)$ has as its solution $\Phi=\pm m/\sqrt{\alpha}$.

Fig. \ref{fig3} shows the behavior of the invariant cosmological acceleration $\Omega$
\begin{equation}\label{Omega}
\Omega=\frac{\ddot{a}a}{a^2}\equiv 1+\frac{\dot{H}}{H^2}
\end{equation}
near the initial singularity. As in the case of the asymmetric scalar doublet \cite{Yu_Kokh_24_2} near the singularity $\Omega\to-1$, which corresponds to the extremely rigid equation of state near the cosmological singularity.

\TwoFigs{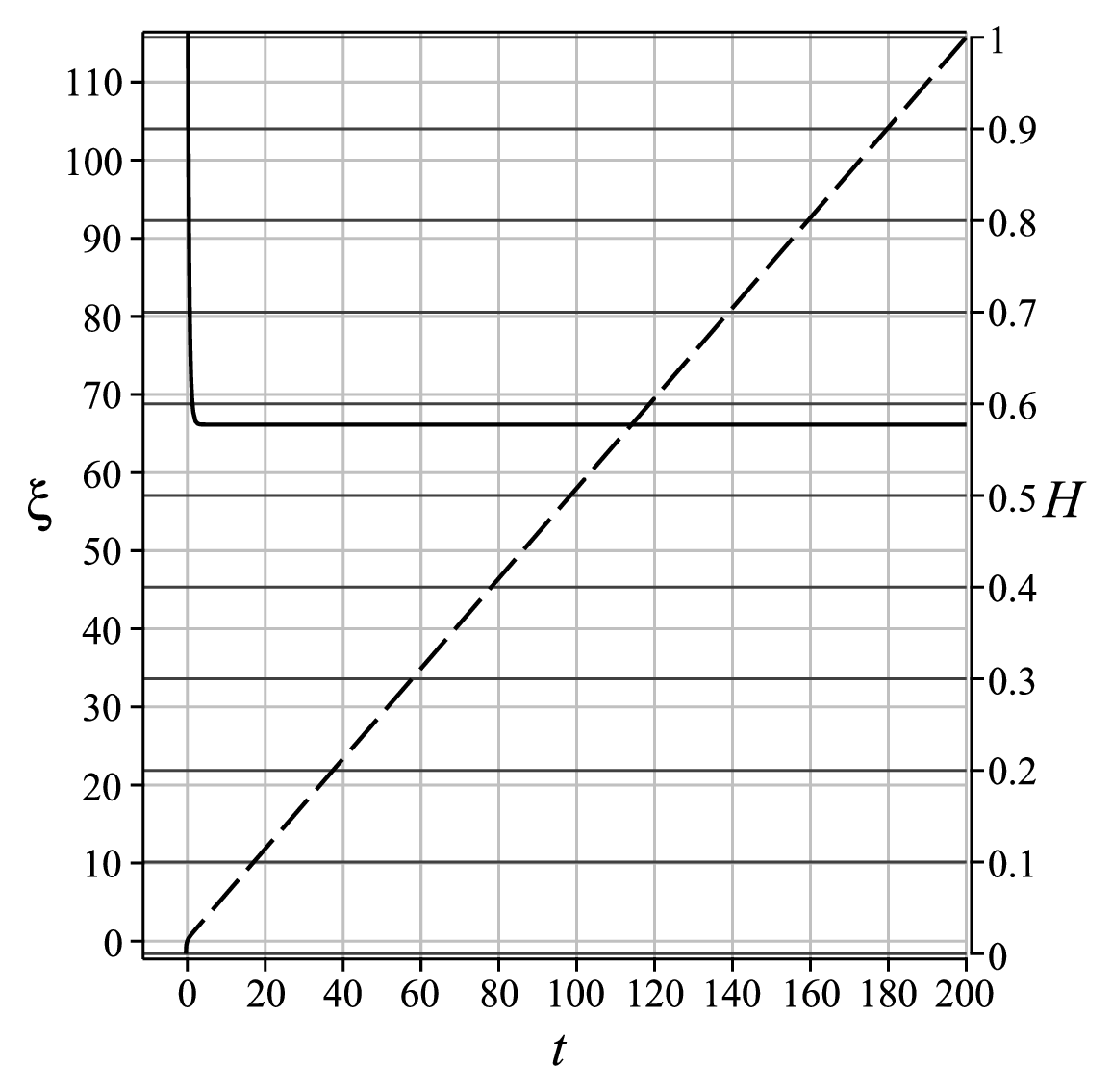}{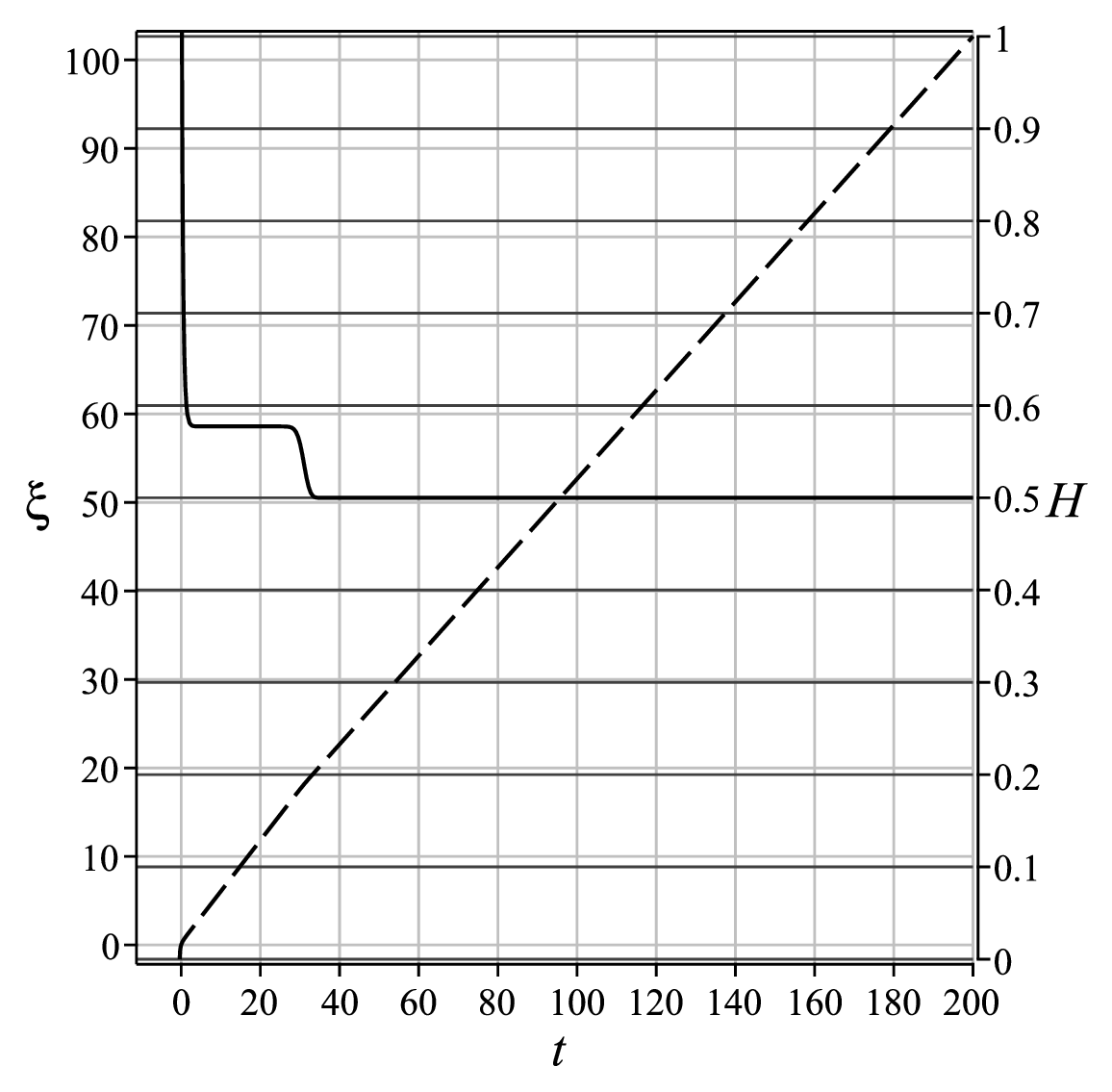}{\label{fig1}Evolution of scale functions $\xi(t)$ (dashed line) and $H(t)$ (solid line) in the model $\mathbf{ S^{(0)}}$: $\mathbf{P}=\mathbf{P^{(0)}_0}$; $\mathbf{I}=\mathbf{I_0}$.}{\label{fig2}Evolution of the scale functions $\xi(t)$ (dashed line) and $H(t)$ (solid line) in the model $\ mathbf{S}$: $\mathbf{P}=\mathbf{P_0}$; $\mathbf{I}=\mathbf{I_0}$.}
\TwoFigs{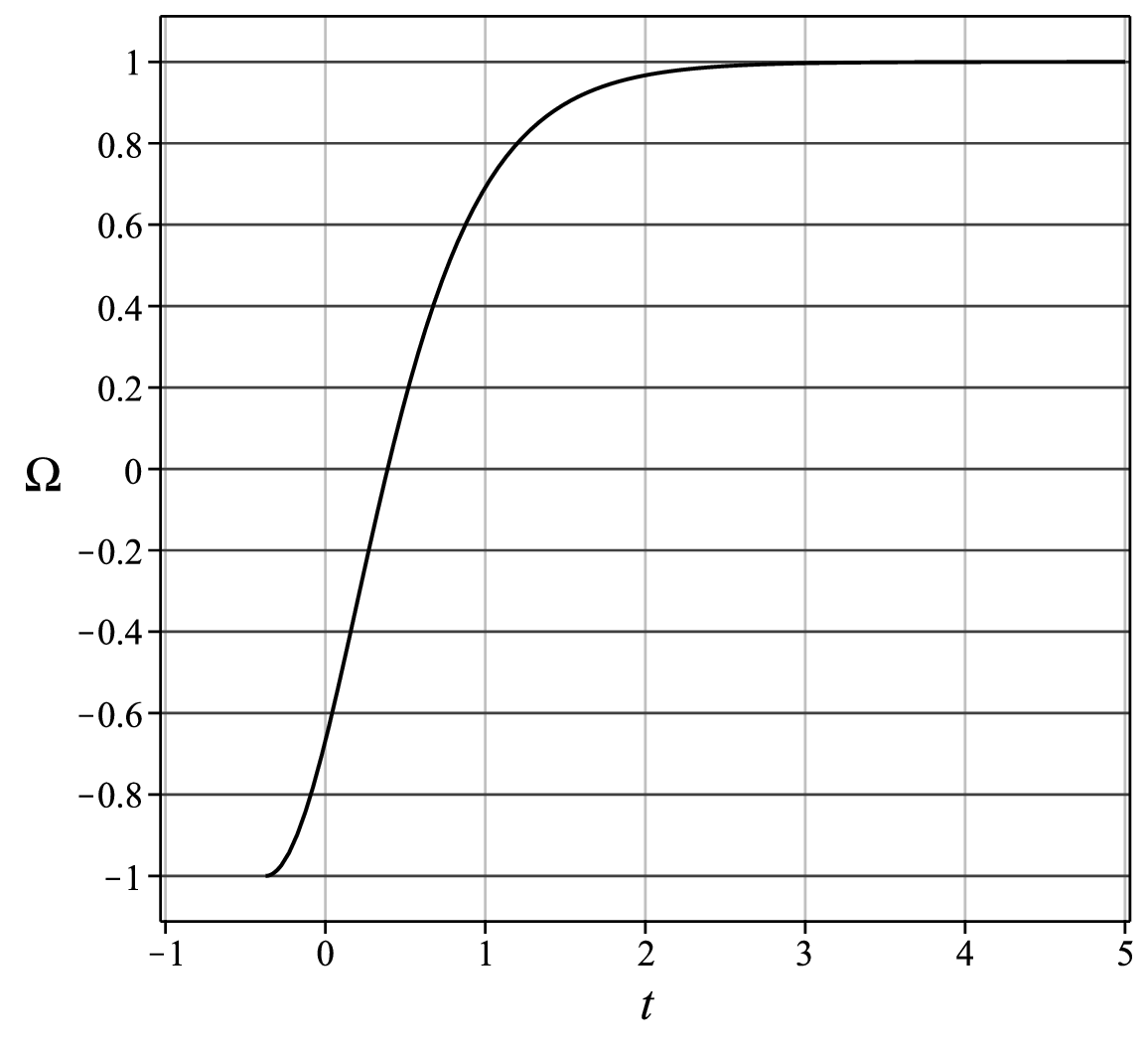}{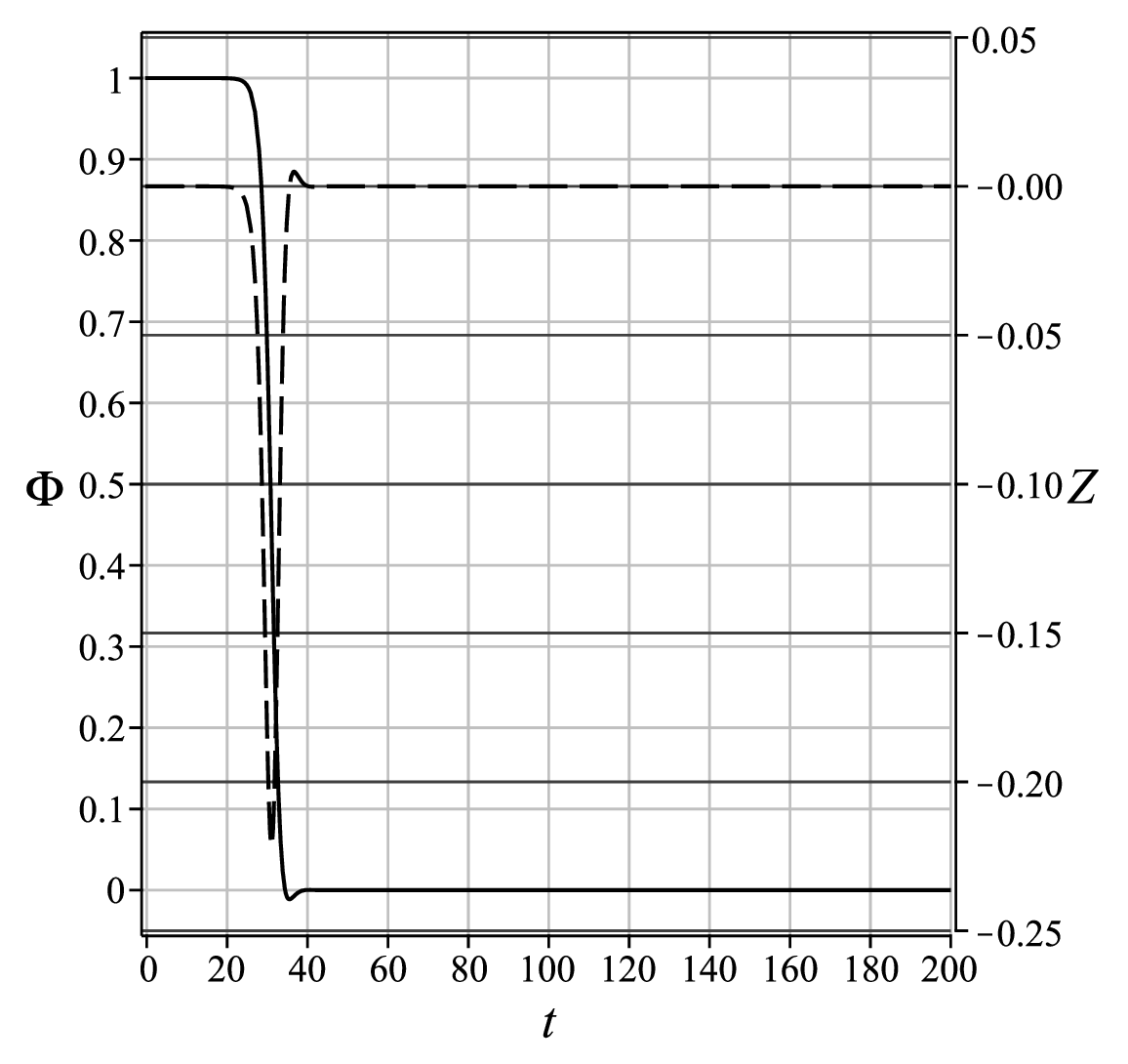}{\label{fig3}Evolution of the cosmological acceleration $\Omega(t)$ near the cosmological singularity $t =t_0$ in the $\mathbf{S}$ model: $\mathbf{P}=\mathbf{P_0}$; $\mathbf{I}=\mathbf{I_0}$.}{\label{fig4}Evolution of the potential $\Phi(t)$ (solid line) and its derivative $Z(t)$ (dashed line) in the $\mathbf{S}$ model: $\mathbf{P}=\mathbf{P_0}$; $\mathbf{I}=\mathbf{I_0}$.}
In Fig. \ref{fig4} one can observe the process of evolution of the scalar potential and its derivative from an unstable state to a stable one. In this case, in the model $\mathbf{S^{ (0)}}$ the values of the potential and its derivative $\Phi=1,Z=0$ are preserved. For this reason, we do not present the graphs of these quantities here.

\subsection{Initial conditions outside singular points}
For the initial state above the saddle point $M^+_+$ ($\Phi_0>1$), cosmological models do not have an infinite future and are short-lived even at very small excesses of the critical potential value, and therefore are of no physical interest. For example, for $\Phi_0=1.01$ and other identical conditions, the cosmological history of the model ends with the Big Rip already at the time $t_{br}\approx 6$ on the Planck time scale.

For the initial state below the saddle point $\Phi_0<1$, we still obtain models with an initial singularity $t_0$ and an infinite inflationary future, and even in the case of a neutral fluid, this late inflation corresponds to a stable point of the dynamic system. In Fig. \ref{fig5} -- \ref{fig6} shows the evolution of the scale functions $\xi(t)$ and $H(t)$ of the $\mathbf{S^{(0)}}$ and $\mathbf{S}$ models for initial conditions with a potential slightly below the saddle point ($\Phi_0=0.99$).
\begin{equation}\label{I_1}
\mathbf{I_1}=[0.99,0,0.2,1].
\end{equation}
Both models have an initial singularity at time $t_0\approx-.37449208$ and an infinite inflationary future with the Hubble parameter $H=0.5$. At the early stages, the models have a small ``shelf'' of early inflation corresponding to the saddle point. At this point, as can be seen from the graphs, the models become practically indistinguishable.
\TwoFigsReg{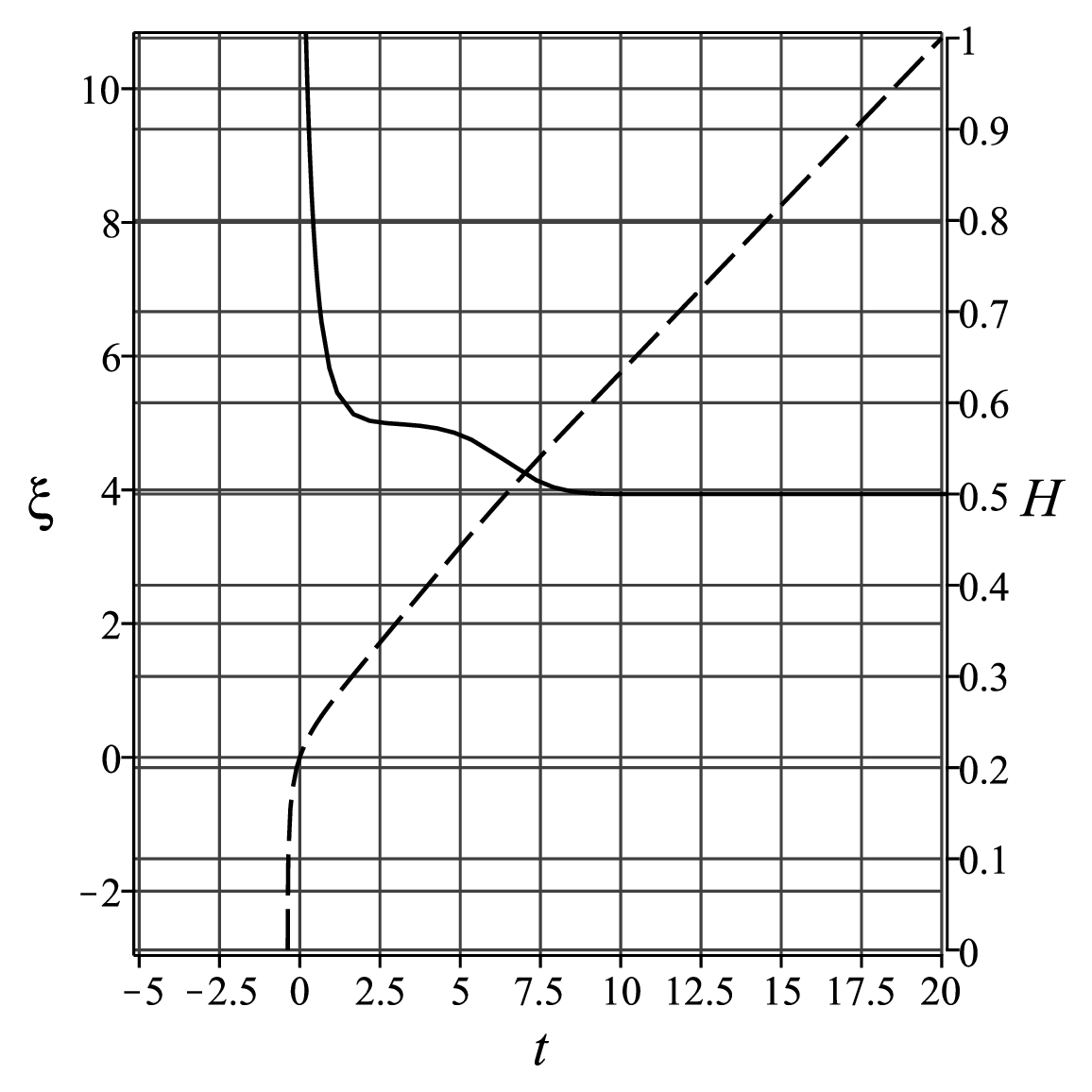}{6}{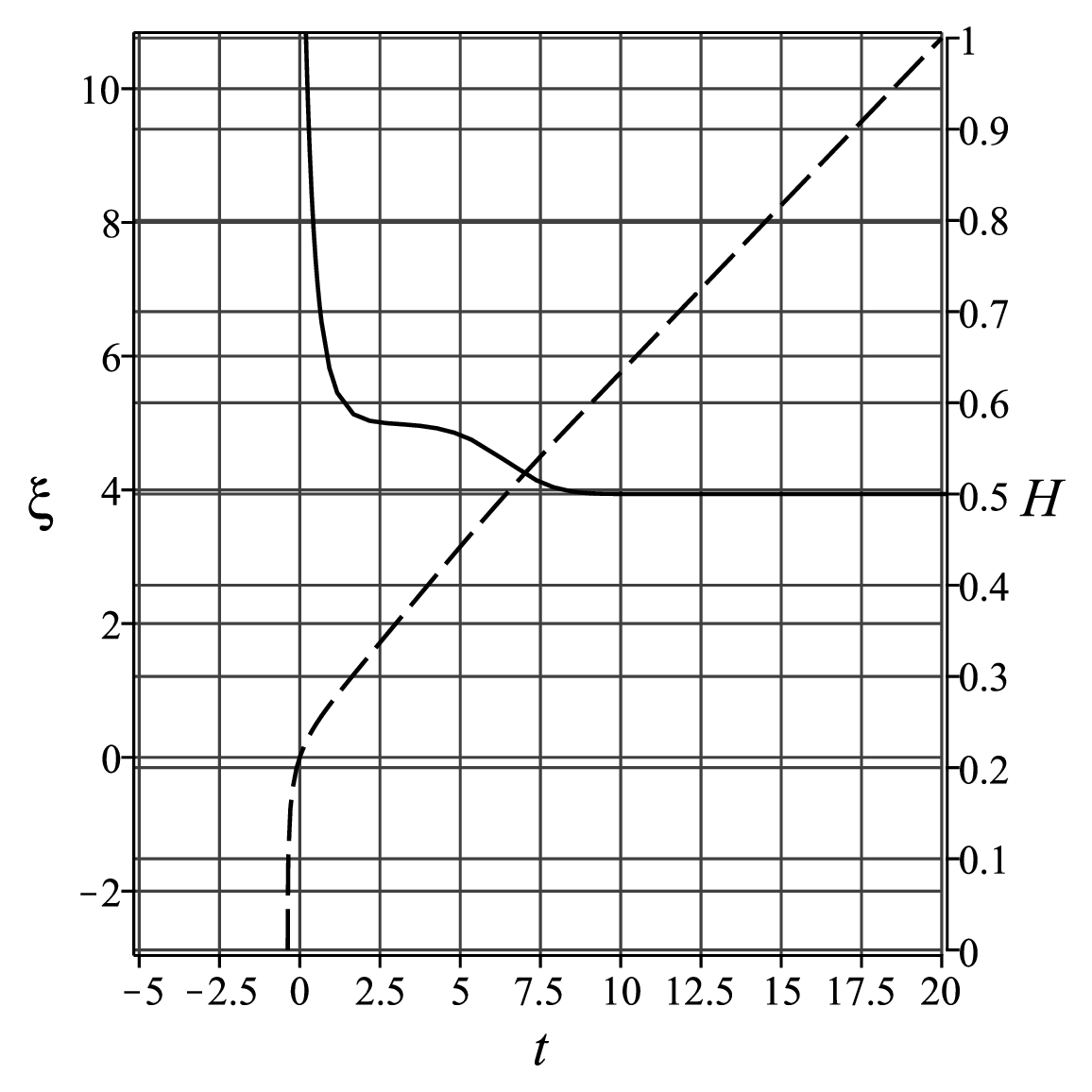}{6}{\label{fig5}Evolution of scale functions $\xi(t)$ (dashed line) and $H(t)$ (solid line) in the $\mathbf{S^{(0)}}$ model: $\mathbf{P}=\mathbf{P^{(0)}_0}$; $\mathbf{I}=\mathbf{I_1}$.}{\label{fig6}Evolution of the scale functions $\xi(t)$ (dashed line) and $H(t)$ (solid line) in the model $\ mathbf{S}$: $\mathbf{P}=\mathbf{P_0}$; $\mathbf{I}=\mathbf{I_1}$.} %

\TwoFigsReg{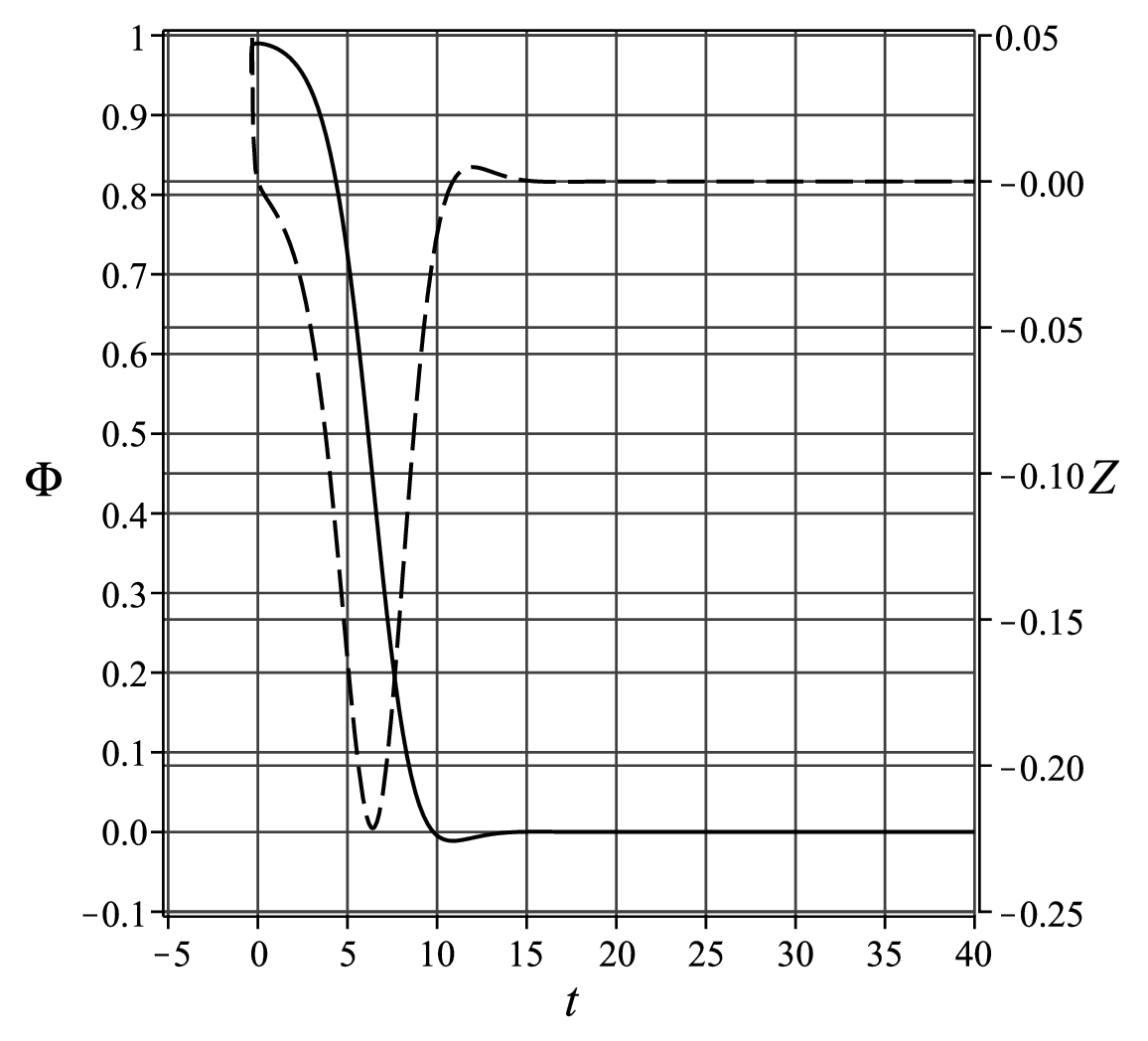}{6}{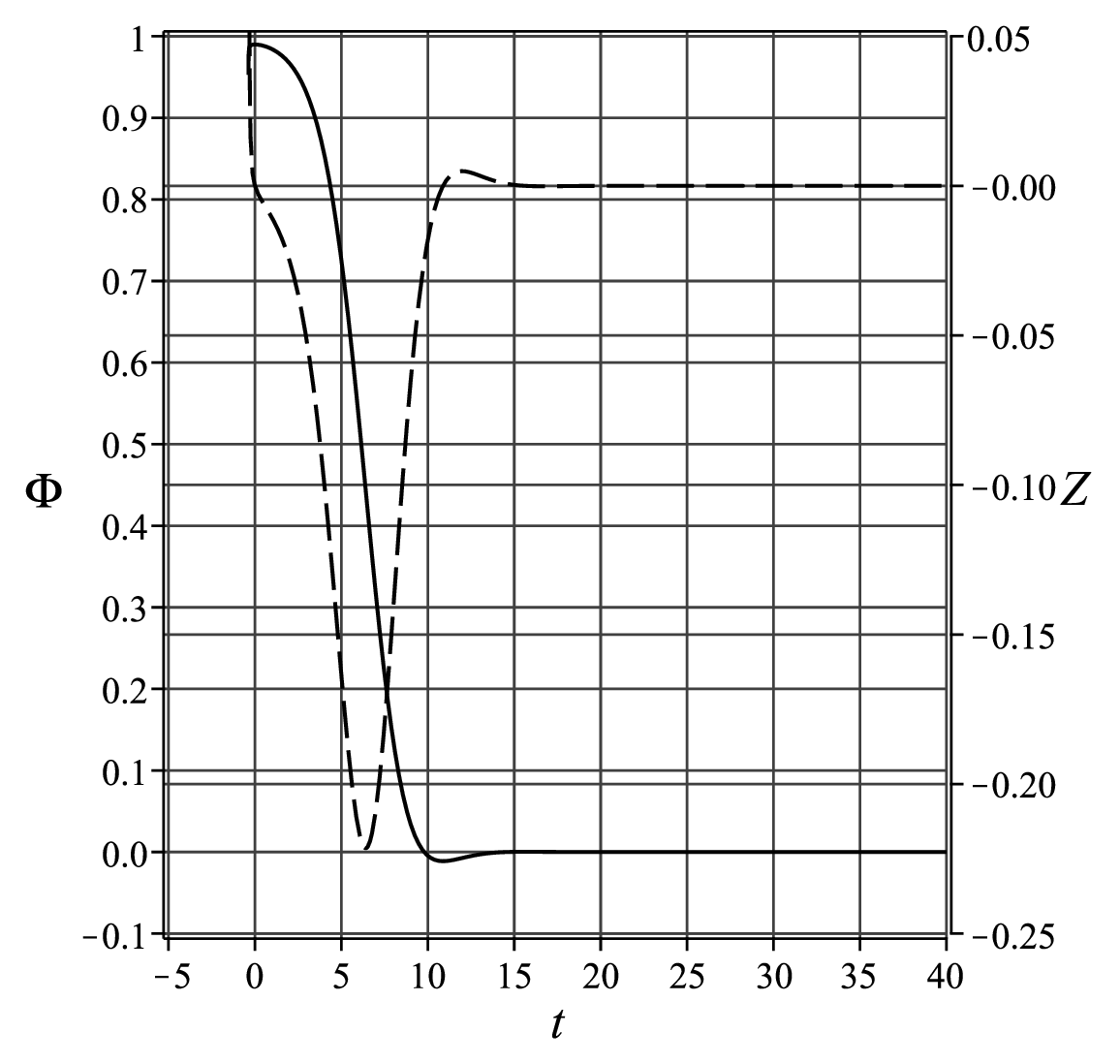}{6}{\label{fig7}Evolution of the scalar potential $\Phi(t)$ (solid line) and its derivative $Z(t)$ (dashed line line) in the $\mathbf{S^{(0)}}$ model: $\mathbf{P}=\mathbf{P^{(0)}_0}$; $\mathbf{I}=\mathbf{I_1}$.}{\label{fig8}Evolution of the scalar potential $\Phi(t)$ (solid line) and its derivative $Z(t)$ (dashed line) in the model $\mathbf{S}$: $\mathbf{P}=\mathbf{P_0}$; $\mathbf{I}=\mathbf{I_1}$.}

Fig. \ref{fig9} -- \ref{fig10} shows the evolution of the scaling functions $\xi(t)$ and $H(t)$ of the models $\mathbf{S^{(0)}_0}$ and $\mathbf{S_0}$ for initial conditions with an even smaller value of the initial potential ($\Phi_0=0.1$).
\begin{equation}\label{I_2 }
\mathbf{I_2}=[0.1,0,0.2,1].
\end{equation}
Both models here also have an initial singularity at time $t_0\approx-0.37447732$ and an infinite inflationary future with the Hubble parameter $H=0.5$, but they lose the ``shelf'' of early inflation corresponding to the saddle point.
\TwoFigsReg{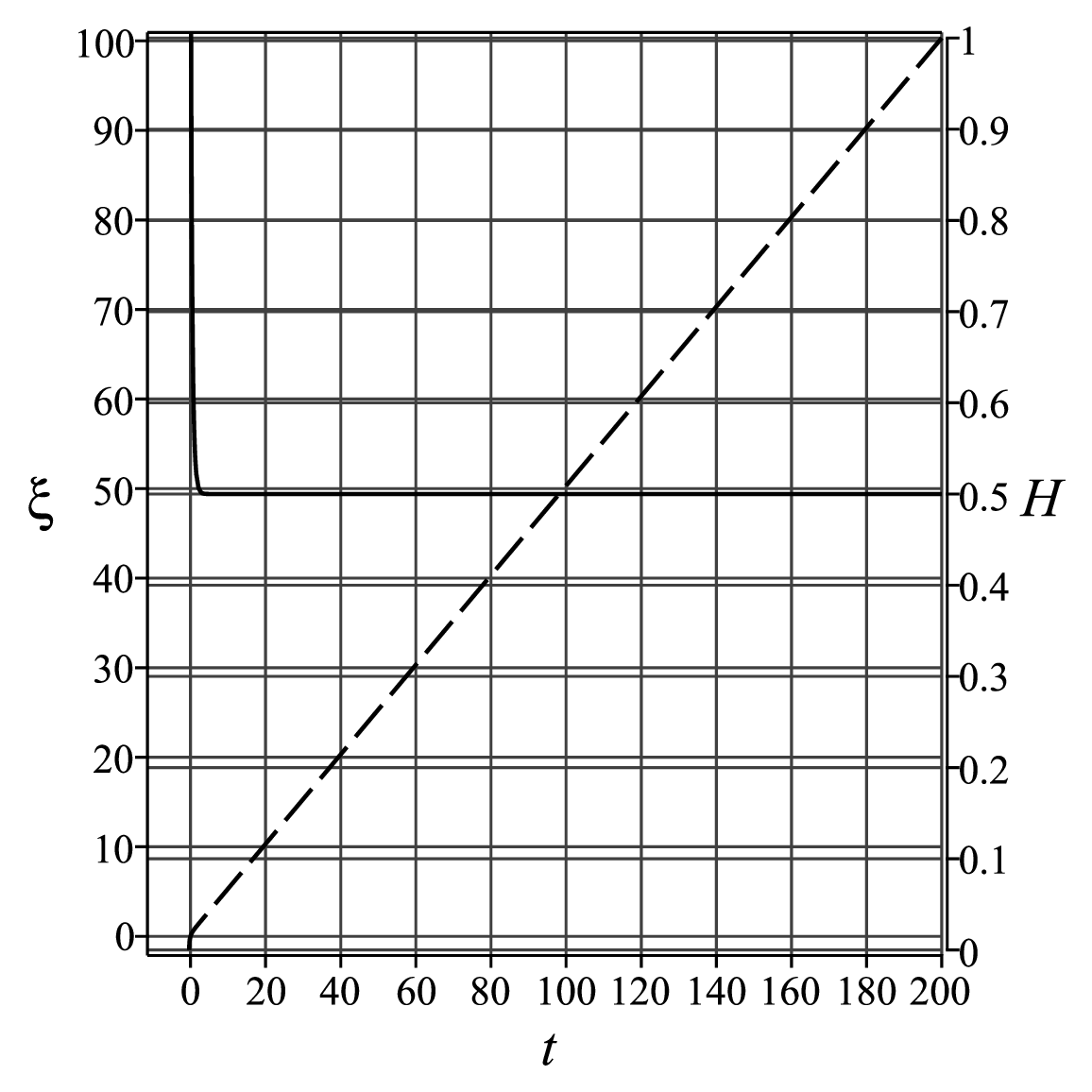}{6}{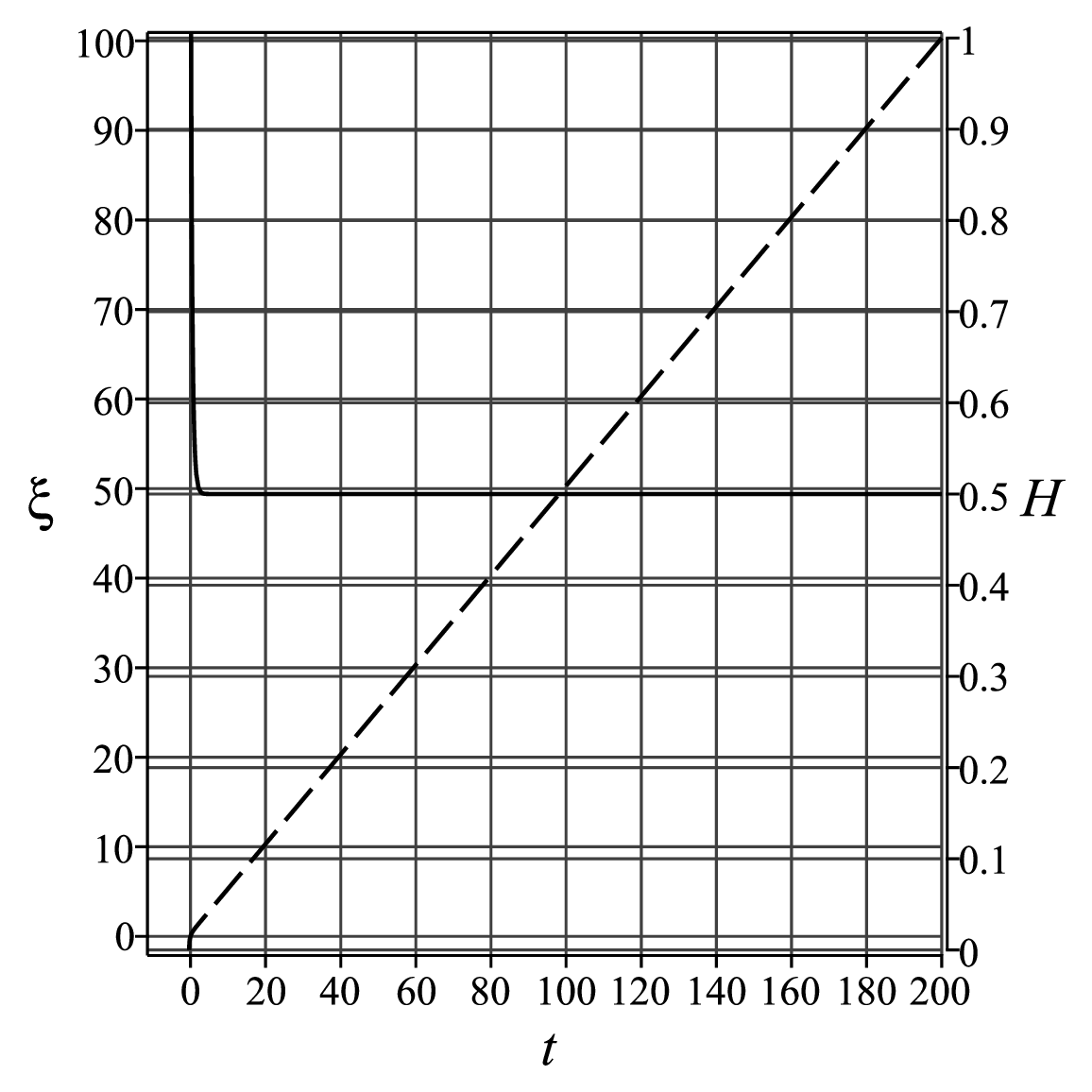}{6}{\label{fig9}Evolution of the scale functions $\xi(t)$ (dashed line) and $H(t)$ (solid line) in the model $\mathbf{S^{(0)}}$: $\mathbf{P}=\mathbf{P^{(0)}_0}$; $\mathbf{I}=\mathbf{I_1}$.}{\label{fig10}Evolution of the scale functions $\xi(t)$ (dashed line) and $H(t)$ (solid line) in the model $\mathbf{S}$: $\mathbf{P}=\mathbf{P_0}$; $\mathbf{I}=\mathbf{I_1}$.}

Fig. \ref{fig11} -- \ref{fig12} shows the graphs of the evolution of scalar potentials and their derivatives in these models.
\TwoFigsReg{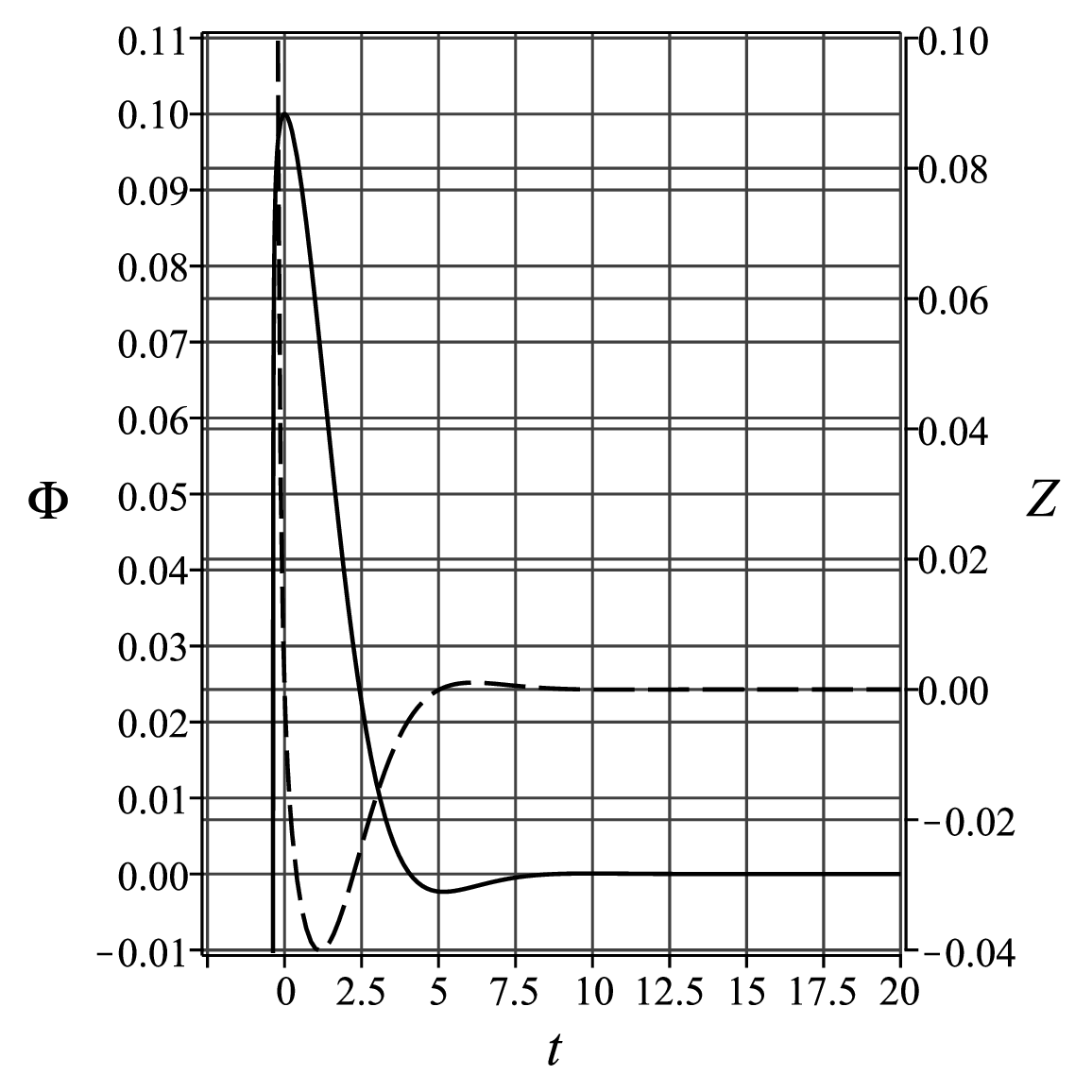}{6}{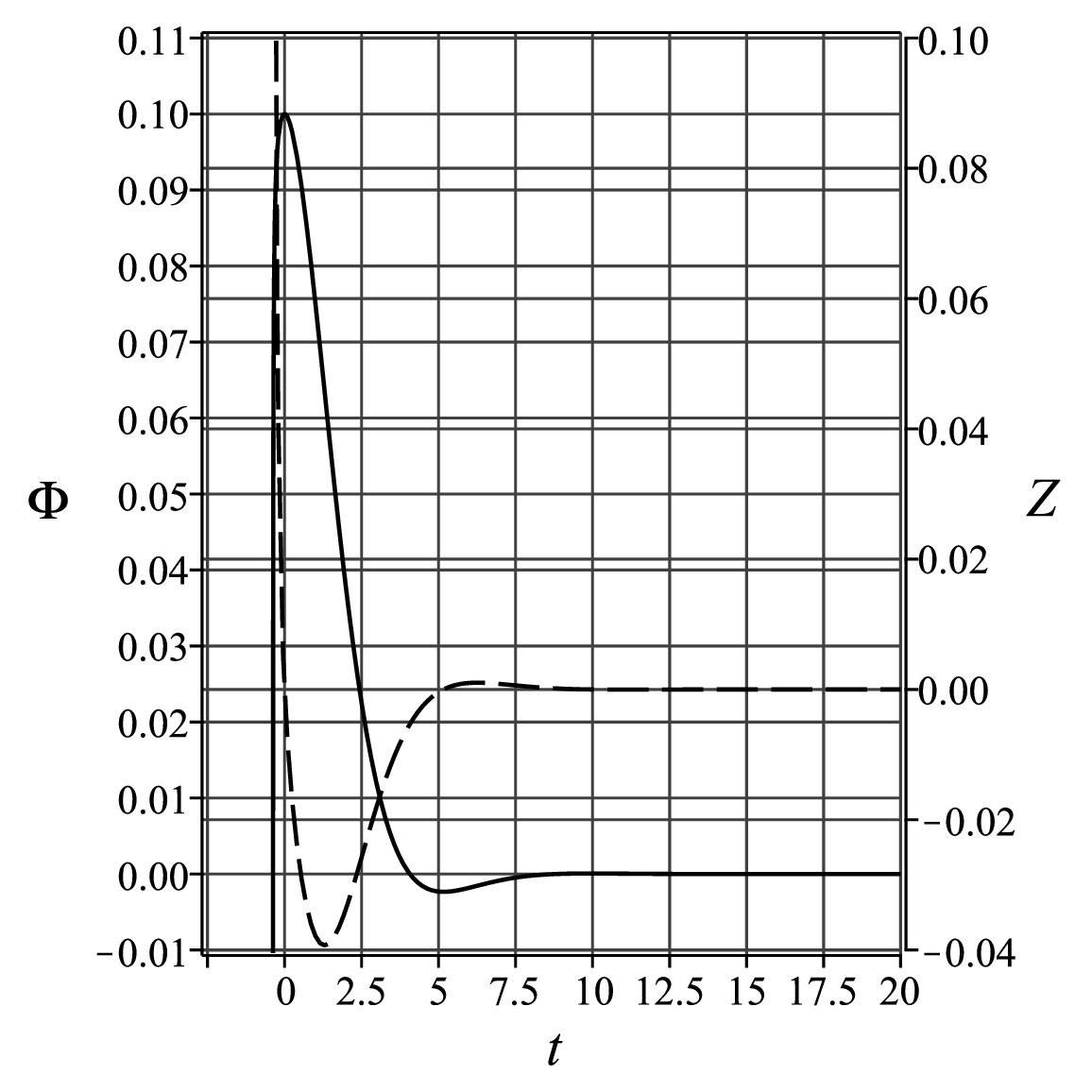}{6}{\label{fig11}Evolution of the scalar potential $\Phi(t)$ (solid line) and its derivative $Z(t)$ (dashed line) in the model $\mathbf{S^{(0)}}$: $\mathbf{P}=\mathbf{P^{(0)}_0}$; $\mathbf{I}=\mathbf{I_2}$.}{\label{fig12}Evolution of the scalar potential $\Phi(t)$ (solid line) and its derivative $Z(t)$ (dashed line) in the $\mathbf{S}$ model: $\mathbf{P}=\mathbf{P_0}$; $\mathbf{I}=\mathbf{I_2}$.}

Thus, to summarize the results of this section, we note the following:
\begin{enumerate}\label{Resume_Fon}
\item The background models $\mathbf{S}$ and $\mathbf{S^{(0)}}$ in the case of small values ??of the scalar charge differ significantly only at the saddle points of the model. These models have an initial singularity corresponding to the total
ultimately rigid equation of state, and inflation at late stages. At the saddle points, the models have different Hubble parameters at the late stage of inflation.
\item For $|\Phi_0|>|\Phi_\pm|\equiv m/\sqrt{\alpha}$ the models also have a final singularity (Big Rip), and for very small excesses of the potential $|\Phi_\pm|$ the cosmological lifetime of the models is reduced to
several Planck times.
\item For $|\Phi_0|<|\Phi_\pm|$ the behavior of the $\mathbf{S}$ and $\mathbf{S^{(0)}}$ models becomes practically indistinguishable. During the transition of the model to the inflation regime the scalar field tends to zero, and the dynamic system
passes into a state of stable equilibrium.
\end{enumerate}
\section{Linear flat perturbations of the cosmological model}
\subsection{Flat perturbations}
We write the metric with gravitational perturbations in the form (see, for example, \cite{Land_Field}):
\begin{eqnarray}
\label{metric_pert}
ds^2=ds^2_0-a^2(\eta)h_{\alpha\beta}dx^\alpha dx^\beta,
\end{eqnarray}
where
\begin{equation}\label{ds_00}
ds_0^2=a^2(\eta)(d\eta^2-dx^2-dy^2-dz^2)
\end{equation}
-- unperturbed Friedmann metric \eqref{ds_0} in terms of conformal time coordinate $\eta$ such that:
\begin{equation}\label{t=a_eta}
t=\int a(\eta)d\eta.
\end{equation}
The transition to a conformal time variable $\eta$ is necessary, firstly, because the functions $\exp(i\mathbf{nr})$ are eigenfunctions of the Laplace operator precisely on the metric \eqref{ds_00}, but not on the metric \eqref{ds_0}, and, secondly, to bring our model into line with the standard model of Friedmann's world perturbation theory. In doing so, we will have to recalculate the formulas of the previous sections for the new time variable. The question arises: why didn't we do this from the very beginning? The fact is that the unperturbed (background) system of Einstein equations -- a scalar field in terms of the time variable $\eta$, unlike the dynamic system $\mathbf{S_4}$, clearly depends on the scale factor $a(\eta)$ and is therefore not autonomous. In this case, the 5-dimensional dynamic system is autonomous, which significantly complicates its analysis and numerical integration (see in this connection \cite{TMF_21}). Therefore, it is simpler to describe the background state in terms of cosmological time $t$, and the perturbed quantities in terms of the time variable $\eta$, and then recalculate them for cosmological time.

We draw attention to the conformal factor $-a^2(\eta)$ before the covariant amplitudes of the disturbances, which disappears for mixed components of the disturbances $h^\alpha_\beta$. In this case, the covariant disturbances of the metric are equal to:
\begin{equation}\label{dg}
\delta g_{\alpha\beta}=-a^2(t)h_{\alpha\beta}.
\end{equation}
Further:
\begin{eqnarray}\label{defh1}
 h^\alpha_\beta=h_{\gamma\beta}g^{\alpha\gamma}_0\equiv-\frac{1}{a^2}h_{\alpha\beta};\\
 \label{defh2}
h\equiv h^\alpha_\alpha\equiv  g^{\alpha\beta}_0h_{\alpha\beta}=
 -\frac{1}{a^2}(h_{11}+h_{22}+h_{33}).
\end{eqnarray}

In what follows we will consider only longitudinal perturbations of the metric, keeping in mind the problem of gravitational stability of plane perturbations, for definiteness directing the wave vector along the $Oz$ axis. In this coordinate system
\begin{eqnarray}\label{nz1}
 h_{11}=h_{22} =\frac{1}{3}[\lambda(t)+\frac{1}{3}\mu(t)]\mathrm{e}^{inz};\;
\label{nz13}
h=\mu(t)\mathrm{e}^{inz};\; h_{12}=h_{13}= h_{23}=0;\;
\label{nz2}
h_{33}=\frac{1}{3}[-2\lambda(t)+\mu(t)]\mathrm{e}^{inz}.
\end{eqnarray}

Assuming further for the perturbations of the scalar field, the energy density and the fluid velocity vector
\begin{eqnarray}\label{delta_Phi}
\Phi(z,\eta)=\Phi_0(\eta)+\phi(\eta) \mathrm{e}^{inz}; & \displaystyle u^i=\frac{1}{a(\eta)}\delta^i_4+\frac{\upsilon(\eta)}{a(\eta)}\mathrm{e}^{inz}\delta^i_3; & \displaystyle
\varepsilon(z,\eta)=\varepsilon_0(\eta)+\delta\varepsilon(\eta)\mathrm{e}^{inz},
\end{eqnarray}
where $\Phi_0(\eta)$, $\varepsilon_0(\eta)$ (as well as $p_0(\eta)$ and $\sigma_0(\eta)$) are unperturbed (background) values ??of the corresponding quantities considered in the previous section, and $\phi(\eta)$, $\upsilon(\eta)$ and $\delta\varepsilon(\eta)$ are the amplitudes of their small perturbations. Using the formulas \eqref{sigma0=}, \eqref{p=p(rho)}, we obtain expressions for the perturbations of the scalar charge density and pressure\footnote{In what follows, to simplify notation, we will omit the argument $\eta$ in functions and denote derivatives with respect to this argument using a prime, -- $\phi'$.}:
\begin{eqnarray}\label{d_sigma}
\sigma(z,\eta)=\sigma_0+\delta\sigma\mathrm{e}^{inz}; & \sigma_0= e^2\Phi_0\varepsilon_0; & \delta\sigma= e^2(\phi\varepsilon_0+\Phi_0\delta\varepsilon)\mathrm{e}^{inz};\nonumber\\
\label{d_p}
p(z,\eta)=p_0(\eta)+\delta p(\eta)\mathrm{e}^{inz}; & \displaystyle p_0=\frac{1}{3}\varepsilon_0(1-e^2\Phi^2_0); & \displaystyle \delta p=\frac{1}{3}[\ \delta\varepsilon(1-e^2\Phi^2_0)-2e^2\varepsilon_0\Phi_0\phi\ ]\mathrm{e}^{inz};\nonumber\\
\delta\varepsilon+\delta p= & \displaystyle \frac{1}{3}[\ \delta\varepsilon(4-e^2\Phi^2_0)-2e^2\varepsilon_0\Phi_0\phi\ ]\mathrm{e}^{inz}. &
\end{eqnarray}
\subsection{Equations for model perturbations}
\subsubsection{Unperturbed state in terms of time variable $\eta$}
Since we will need Einstein's background equations, the scalar field equation, and the conservation laws of the total EMT to obtain the equations for perturbations, we will write them in terms of the variable $\eta$.\\
Non-trivial Einstein's background equations:
\begin{eqnarray}\label{11-44}
\frac{1}{2}(EQ^1_1-EQ^4_4): & \displaystyle \frac{1}{a}\left(\frac{a''}{a^2}+2\frac{a'^2}{a^3}\right)+\frac{1}{a^2}{\Phi'}_0^2+4\pi(\varepsilon_0+p_0)=0;\\ \label{44} EQ^4_4: & \displaystyle 3\frac{a'^2}{a^4}-\ frac{1}{2}\frac{{\Phi'}_0^2}{a^2}-8\pi\varepsilon_0 +\frac{\alpha}{4}\left(\Phi_0^2-\frac{m^2}{\alpha}\right)^2-\Lambda_0=0.
\end{eqnarray}
Note that the first term on the left-hand side of equation \eqref{11-44} is $\dot{H}\equiv H'/a$:
\begin{equation}\label{dH(eta)}
H(\eta)=\frac{a'}{a^2};\quad H'=\frac{a''}{a^2}+2\frac{a'^2}{a^3}.
\end{equation}
Background equation of the scalar field:
\begin{eqnarray}\label{Eq_S_eta}
{\Phi_0}''+2\frac{a'}{a}\phi'+a^2\Phi_0(m^2-\alpha\Phi^2_0)=-8\pi a^2\sigma_0,
\end{eqnarray}
Non-trivial background equation of conservation of the total EMT --
\begin{eqnarray}\label{T^k_3,k}
{\varepsilon_0}'+3\frac{a'}{a}(\varepsilon_0+p_0)-{\Phi_0}'\sigma_0=0.
\end{eqnarray}
In the obtained equations it is necessary to substitute unperturbed values ??of macroscopic scalars $p_0$ and $\sigma_0$ from \eqref{d_sigma}.

\subsubsection{Perturbed equation of a scalar field}
Expanding the equations of a scalar field in terms of the smallness of the perturbations, we obtain in the linear approximation the perturbed equation of the field
\begin{eqnarray}\label{Eq_dPhi}
\phi''+2\frac{a'}{a}\phi'+\bigl[n^2+a^2(m^2-3\alpha\Phi^2_0)\bigr]\phi+\frac{1}{2}\mu'\Phi'_0=-8\pi a^2e^2(\delta\varepsilon\Phi_0+\varepsilon_0\phi).
\end{eqnarray}
\subsubsection{Perturbations of the total energy-momentum tensor}
Thus, non-zero perturbations of the components of the total energy-momentum tensor have the form:
\begin{eqnarray}\label{TS^a_a}
8\pi\mathrm{e}^{-inz}\delta(S^\beta_\beta+T^\beta_\beta)=-\frac{1}{a^2}\Phi'_0\phi'+\Phi_0\phi(m^2-\alpha\Phi^2_0)-8\pi\delta p, & (\beta=\overline{1,3});\\
\label{FS^4_4}
8\pi\mathrm{e}^{-inz}\delta(S^4_4+T^4_4)=\frac{1}{a^2}\Phi'_0\phi'+\Phi_0\phi(m^2-\alpha\Phi^2_0)-8\pi\delta \varepsilon; &\\
\label{TS^3_4}
8\pi\mathrm{e}^{-inz}\delta(S^3_4+T^3_4)=-8\pi\mathrm{e}^{-inz}\delta(S^4_3+T^4_3)=-\frac{i}{a^2}\Phi'_0\phi'+\upsilon(\varepsilon_0+p_0), &
\end{eqnarray}
where it is necessary to substitute the expression for the pressure perturbation $\delta p$ from \eqref{d_p}.

\subsubsection{Perturbed conservation equations}
These equations are obtained from the conservation laws of the total EMT
\begin{equation}
\nabla_i (S^i_k+T^i_k)=0.
\end{equation}
Non-trivial perturbed conservation equations have the form:
\begin{eqnarray}\label{dT^k_3,k}
k=3: & \displaystyle in(\phi\sigma_0+\delta p)a^4+\bigl[\upsilon(\varepsilon_0+p_0)a^4\bigr]'=0;\\
\label{dT^k_4,k}
k=4: & \displaystyle \delta\varepsilon'+3\frac{a'}{a}(\delta\varepsilon+\delta p)+ in(\varepsilon_0+p_0)\upsilon +\frac{1}{2}\mu'\biggl(\varepsilon_0+p_0+\frac{{{\Phi_0}'}^2}{8\pi a^2}\biggr)-{\Phi_0}'\delta\sigma-\sigma_0\phi'.
\end{eqnarray}
\subsubsection{Perturbed Einstein equations\label{flat_pert_Einst}}
Non-trivial perturbed Einstein equations have the form:
\begin{eqnarray}\label{d11-d33}
\delta(EQ^1_1-EQ^3_3): & \displaystyle -\frac{1}{a^2}(a^2\lambda')'+\frac{1}{3}(\lambda+\mu)n^2=0;\\
\label{d33}
\delta EQ^3_3: & \displaystyle {\Phi_0}'\phi'-a^2\Phi_0\phi(m^2-\alpha\Phi^2_0)+8\pi a^2\delta p +\frac{1}{3a^2}(a^2(\lambda+\mu)')'=0;\\
\label{d44}
\delta EQ^4_4: & \displaystyle -{\Phi_0}'\phi'+\frac{a'}{a}\mu'+\frac{1}{3}(\lambda+\mu)n^2-a^2\Phi_0\phi(m^2-\alpha\Phi^2_0)-8\pi a^2\delta\varepsilon=0;\\
\label{d43}
\delta EQ^3_4: & \displaystyle 8\pi a^2(\varepsilon_0+p_0)\upsilon=in\bigl(\phi{\Phi_0}'+\frac{1}{3}(\lambda+\mu)'\bigr).
\end{eqnarray}

Note, first, that the equation for the perturbations of the scalar field \eqref{Eq_dPhi}, as well as the perturbed conservation equations \eqref{dT^k_3,k} -- \eqref{dT^k_4,k} are obtained as differential-algebraic consequences of the four perturbed Einstein equations \eqref{d11-d33} -- \eqref{d43}. Therefore, in fact, there are only 4 ordinary linear differential equations \eqref{d11-d33} -- \eqref{d43} in four unknown functions $\lambda(\eta)$, $\mu(\eta)$, $\varepsilon(\eta)$, and $\upsilon(\eta)$.

Secondly, we note that the equation \eqref{d43} for $\varepsilon_0+p_0\not\equiv0$ is in fact the definition of the magnitude of the velocity of matter $\upsilon(\eta)$. Otherwise, in the absence of an ideal fluid, i.e., \emph{in the vacuum-field model }\eqref{vac_mod}, from \eqref{d43} we obtain
\begin{equation}\label{e+p=0}
\varepsilon_0+p_0\equiv0 \Rightarrow \phi{\Phi_0}'+\frac{1}{3}(\lambda+\mu)'=0;\quad \varepsilon_0=0\Rightarrow \delta\varepsilon=0 .
\end{equation}
Using this relation in the remaining Einstein equations \eqref{d11-d33} -- \eqref{d44}, we obtain 3 independent linear differential equations for two unknown functions -- $\phi(\eta)$ and, for example, $\lambda(\eta)$.
The system is essentially overdetermined and, generally speaking, inconsistent. Note that adding the off-diagonal term $h_{43}$ to the metric does not save the situation -- in this case, the number of independent Einstein equations increases.

Thus, the following statement is true.
\begin{stat}\label{epsilon_not0}\hskip -4pt \textbf{.}

Longitudinal perturbations of the Friedmann metric \eqref{nz1} -- \eqref{nz2} in the model with a scalar field are possible only in the presence of an isotropic fluid.
\end{stat}
\subsubsection{Degenerate case of the vacuum-field model:\label{pert_vac} $\mathbf{S^{(00)}}$}
The only exception is the case of the vacuum-field model $\mathbf{S^{(00)}_4}$, when
\[\Phi_0=\mathrm{Cons}\ \Rightarrow (\lambda+\mu)'=0. \]
But then $(\lambda+\mu)=0$, since any constant in this relation is eliminated by an admissible transformation of the metric. Considering that in the absence of an ideal fluid $\delta p=\delta\varepsilon=0$, we obtain from \eqref{d33}
\begin{equation}\label{Phi_0Phi_0=0}
\Phi_0(m^2-\alpha\Phi^2_0)=0.
\end{equation}
In this case, it follows from \eqref{d44} that $\mu=\mathrm{Const}$, and consequently $\lambda=\mathrm{Const}$, i.e., the perturbed metric can be reduced to the unperturbed Friedmann metric by admissible transformations. Thus, in this degenerate case, too, longitudinal perturbations of the metric are absent. In this case, all \emph{linear} perturbations of the Einstein equations \eqref{d11-d33} -- \eqref{d43} and along with them the \emph{linear} perturbations of the total EMT \eqref{TS^a_a} -- \eqref{TS^3_4} vanish regardless of the magnitude of the perturbation of the scalar potential $\phi(\eta)$, with respect to which only one field equation \eqref{Eq_dPhi} remains. Note that in the linear approximation, perturbations of the scalar field do not contribute to the energy-momentum of the cosmological system.

When the condition \eqref{Phi_0Phi_0=0} is satisfied, the unperturbed equations of the gravitational and scalar fields are reduced to a single nontrivial Einstein equation \eqref{44}
 \begin{equation}\label{Eq_a(eta)}
  \Phi_0=0 \Rightarrow \frac{a'^2}{a^4}= \displaystyle\frac{\Lambda}{3};  \qquad
 \displaystyle \Phi_0=\pm \frac{m}{\sqrt{\alpha}} \Rightarrow \frac{a'^2}{a^4}=\frac{\Lambda_0}{3},\quad (\Lambda<\Lambda_0).
 \end{equation}
Thus, we obtain the well-known inflationary solution for the scale factor (see, for example, \cite{Weinberg})
\begin{eqnarray}\label{a(eta)}
a(\eta)=\pm\frac{1}{H_0\eta}\Rightarrow \eta=-\frac{1}{H_0}\mathrm{e}^{-H_0t} & \Rightarrow a(t)=\mathrm{e}^{H_0 t}:\\
\Phi_0=0\Rightarrow H_0= \sqrt{\frac{\Lambda}{3}}; & \displaystyle\Phi_0=\pm \frac{m}{\sqrt{\alpha}}\Rightarrow H_0= \sqrt{\frac{\Lambda_0}{3}}.
\end{eqnarray}
Choosing a negative sign in the solution \eqref{a(eta)}, corresponding to the expansion of the Universe (in this case $\eta\in(-\infty,0]$, i.e., the infinite future $t\to+\infty$ corresponds to $\eta\to-0$), we reduce
the equation for the amplitude of the perturbation $\phi(\eta)$ of the scalar field \eqref{Eq_dPhi} to a Bessel-type equation
\begin{equation}\label{Eq_dPhi_Phi_0}
\phi''-2\frac{\phi'}{\eta}+\left[n^2+\frac{m^2-3\alpha\Phi_0^2}{H_0^2\eta^2}\right]\phi=0,
\end{equation}
which has as its general solution:
\begin{eqnarray}\label{phi(eta)}
\phi(\eta)=(-\eta)^{3/2}[\mathrm{C}_1\mathrm{J}_\gamma(n\eta)+\mathrm{C}_2\mathrm{Y}_\gamma(n\eta)];\qquad
\gamma\equiv \frac{3}{2}\left(1-\frac{4}{9H^2_0}(m^2-3\alpha\Phi_0^2)\right)^{1/2},
\end{eqnarray}
where $\mathrm{J}_\nu(z)$ and $\mathrm{Y}_\nu(z)$ are the Bessel functions of the first and second kind, respectively (see \cite{Lebedev})\footnote{The author hopes that the reader will not be confused by the use of the symbol $\nu$, previously used to denote the perturbation of the gravitational field, as an index of the Bessel functions.}. Thus, for the parameter $\gamma$ of the Bessel functions, we obtain according to \eqref{a(eta)}

\begin{eqnarray}\label{nu}
\nu=\displaystyle \frac{3}{2}\sqrt{1-\frac{4m^2}{3\Lambda}},& \displaystyle  (\Phi_0=0); \qquad  \displaystyle \gamma=\frac{3}{2}\sqrt{1+\frac{8m^2}{3\Lambda_0}}, & \displaystyle  \bigl(\Phi_0=\pm\frac{m}{\sqrt{\alpha}}\bigr) .
\end{eqnarray}

Note that in the case of $\Phi_0=0$, when the condition \eqref{attract_eq} ($3\Lambda-4m^2<0$) is satisfied, the state of the unperturbed vacuum-field model is stable, unlike any other cases. Therefore, just in the case of a stable vacuum-field model, the index $\gamma$ of the Bessel functions in the solution \eqref{phi(eta)} is purely imaginary $\gamma^2<0$. Below we consider two examples of solutions \eqref{phi(eta)} for the values of the problem parameters
$P = [n,\alpha,m,\Phi_0,\Lambda_0,C_1,C_2]$:
\begin{eqnarray}
P_0 = [1,1,1,0,1,1,1];\; P_1 = [1,1,1,1,1,1,1].\nonumber
\end{eqnarray}
In this case, the system with parameters $P_0$ is stable, the system with parameters $P_1$ is unstable, and
\[\Lambda=\frac{3}{4};\; \gamma_0=\frac{i}{2}\sqrt{7};\; \gamma_1=\frac{1}{2}\sqrt{11}.\]
Fig. \ref{fig13} and \ref{fig14} show the graphs of $\phi(\eta)$, the real and imaginary parts of the solutions \eqref{phi(eta)}.
It follows from these graphs that in the case of a stable background solution $|\phi(0)|\to 0$, and in the case of an unstable background solution $\phi(0)\to \infty$, i.e., in the case of an unstable background solution, the flat perturbations of the scalar field also become unstable. It should be remembered that the value of the time variable $\eta\to0$ corresponds to an infinitely large cosmological time $t\to\infty$. Note, however, that the argument of the Bessel functions in the solution \eqref{phi(eta)} is not $\eta$, but $n\eta$. In other words, the instability of flat scalar perturbations occurs for long-wave perturbations:
\begin{equation}\label{n_eta->8}
n\eta\gtrsim -1.
\end{equation}

Fig. \ref{fig15} shows the development of instability of scalar field oscillations in the model
\begin{eqnarray}
P_1^1 = [\mathbf{0.1},1,1,1,1,1,1],\nonumber
\end{eqnarray}
in which the wave number $n=0.1$ is an order of magnitude smaller than in the $P_1$ model, which allows us to see a detailed picture of the development of instability in the region \eqref{n_eta->8}: $n\eta\gtrsim-1\Rightarrow \eta\gtrsim-10$.

\Fig{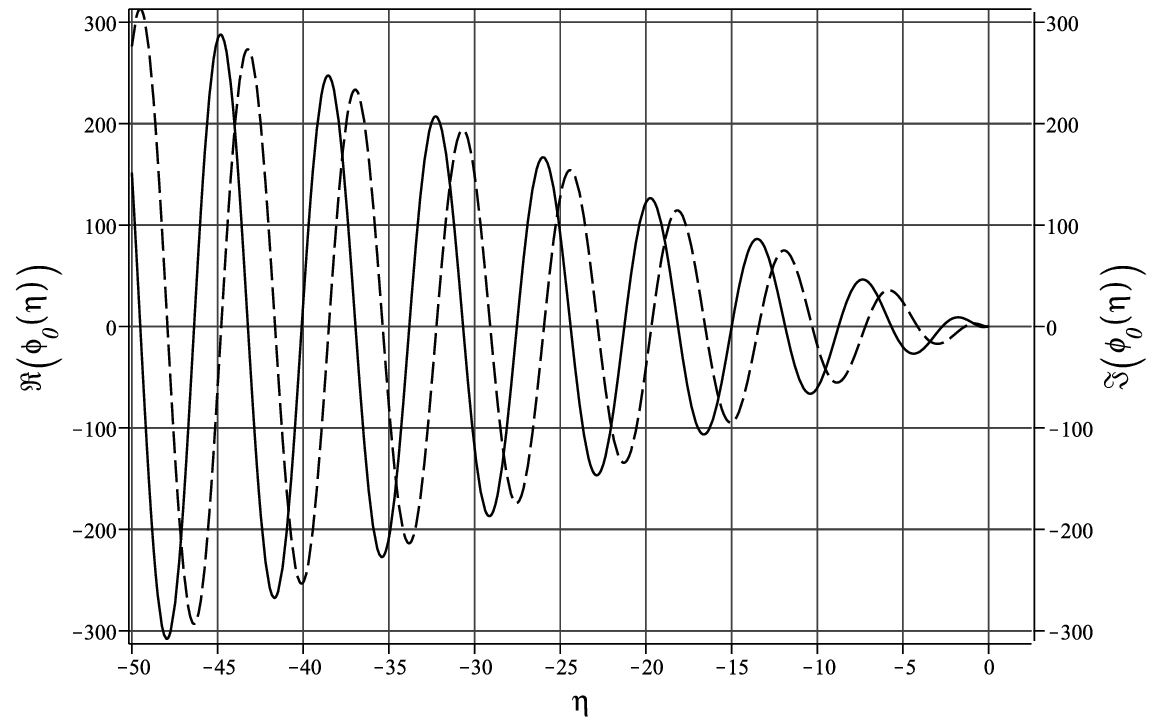}{12}{6}{\label{fig13} Plots of \eqref{phi(eta)} solution in case of stable background state ($P_0$): solid line -- $\Re(\Phi(\eta))$, dashed line -- $\Im(\Phi(\eta))$.}
\vspace{12pt}
\Fig{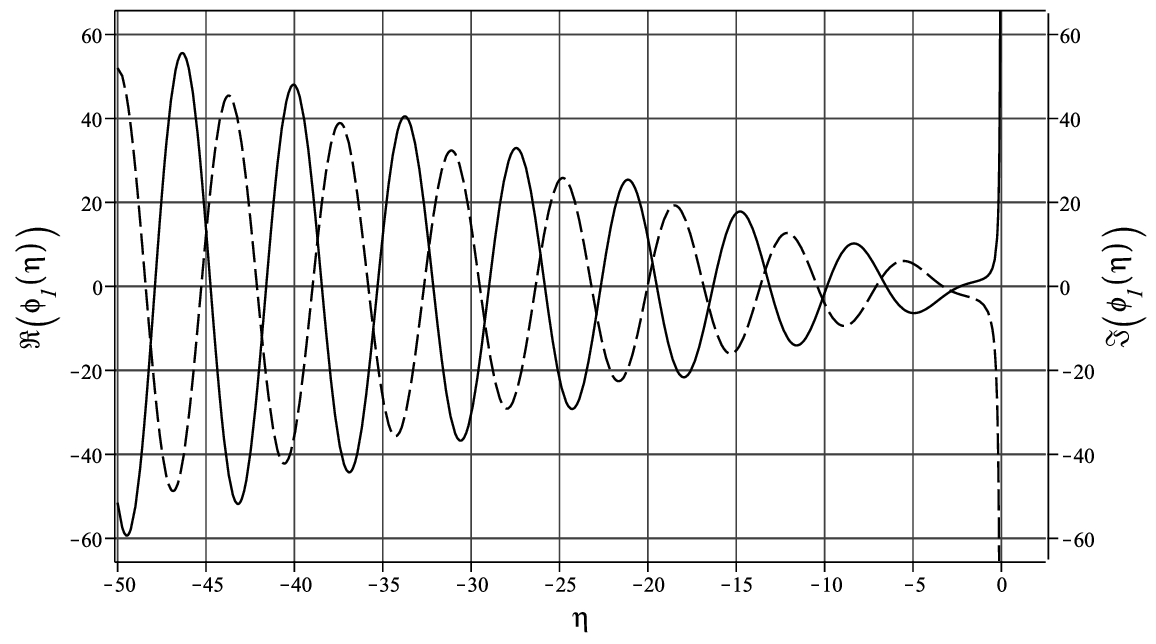}{12}{6}{\label{fig14} Plots of \eqref{phi(eta)} solution in case of unstable background state ($P_1$): solid line -- $\Re(\Phi(\eta))$, dashed line -- $\Im(\Phi(\eta))$.}
\Fig{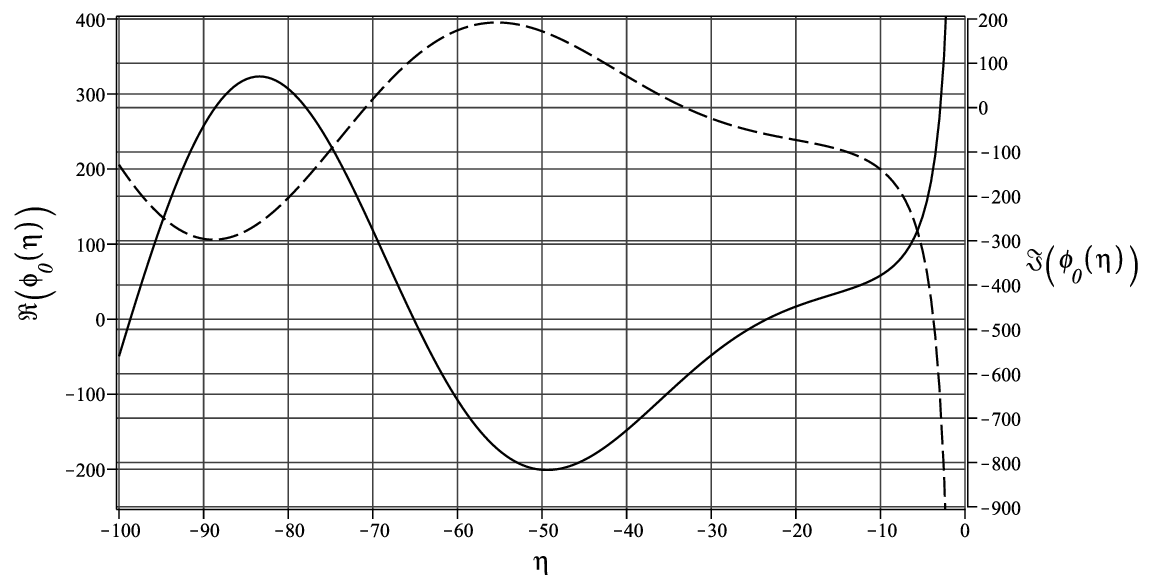}{12}{6}{\label{fig15} Plots of \eqref{phi(eta)} solution in case of unstable background state ($P_1^1$): solid line -- $\Re(\Phi(\eta))$, dashed line -- $\Im(\Phi(\eta))$.}

Thus, the following statement is true.

\begin{stat}\label{epsilon=0}\hskip -4pt \textbf{.}
\begin{enumerate}
\item In the Friedmann vacuum-field model, i.e., in the absence of an isotropic fluid \eqref{vac_mod}, there are no planar longitudinal perturbations of the metric caused by perturbations of the scalar field.
\item In the only case when the background state corresponds to one of the singular points of the model $\mathbf{S^{(00)}}$, oscillations of the scalar field near these singular points (the exact solution \eqref{phi(eta)}) are possible, not
accompanied by perturbations of the Friedmann metric. In this case, in the case of a stable background state, the oscillations of the scalar field decay with time, in the case of an unstable background state, the oscillations of the scalar field become unstable at $n\eta\lesssim1$.
\end{enumerate}
\end{stat}
\section{WKB - approximation}
\subsection{WKB-decomposition}
Let us consider the short-wave approximation in more detail. In \cite{WKB_22} the WKB theory of cosmological evolution of longitudinal gravitational scalar perturbations of the model for a system of scalar charged degenerate fermions in the WKB-approximation is developed:
\begin{eqnarray}\label{WKB}
n\eta\gg 1; & \displaystyle n^2\backsim a^2m^2.
\end{eqnarray}
We represent the perturbation functions $f(\eta)$ in accordance with the WKB method in the form
\begin{equation}\label{Eiconal}
f=\tilde{f}(\eta) \cdot \mathrm{e}^{i\int\!u(\eta)d\eta}; \quad (|u|\sim n \gg \eta^{-1}),
\end{equation}
where $\tilde{f}(\eta)$ and $u(\eta)$ are the perturbation amplitude and eikonal functions that vary slightly along with the scale factor. %
We should keep in mind the following relationships:

\begin{eqnarray}
f'=(\tilde{f}' +iu\tilde{f})\cdot \mathrm{e}^{i\int\!u(\eta)d\eta}; & \displaystyle f''=(\tilde{f}''+2iu\tilde{f}'+iu'\tilde{f}-u^2\tilde{f})\cdot \mathrm{e}^{i\int\!u(\eta)d\eta}
\end{eqnarray}

Substituting the perturbations $\mu(\eta), \lambda(\eta),\phi(\eta),\delta\varepsilon(\eta)$ in the form \eqref{Eiconal} into the perturbed equations of the scalar field \eqref{Eq_dPhi} and Einstein \eqref{d11-d33} -- \eqref{d44} in the first WKB approximation \eqref{Eiconal} leads to the following result:
\begin{eqnarray}
\bigl\{\tilde{\phi}\bigl[n^2-u^2+a^2(m^2-3\alpha\Phi^2_0-8\pi e^2\varepsilon_0)\bigr] -8\pi a^2e^2\Phi_0\tilde{\delta\varepsilon} \bigr\} +i\biggl\{ 2u\tilde{\phi}+u'\tilde{\phi}+2\frac{a'}{a}u\tilde{\phi}+\frac{1}{2}u(\tilde{\nu}-\tilde{\lambda}){\Phi_0}'\biggr\}=0;\nonumber\\
\biggl\{u^2\tilde{\lambda}+\frac{1}{3}n^2\tilde{\nu}\biggr\}-i\biggl\{2\frac{a'}{a}u\tilde{\lambda}+2u\tilde{\lambda}+u'\tilde{\lambda}\biggr\}=0;\nonumber
\end{eqnarray}
\begin{eqnarray}
\biggl\{-\frac{1}{3}u^2\tilde{\nu}-a^2\Phi_0\biggl(m^2-\alpha\Phi_0^2-\frac{16\pi}{3}e^2\varepsilon_0\biggr)\tilde{\phi}+\frac{8\pi}{3} a^2(1-e^2\Phi_0^2)\tilde{\delta\varepsilon} \biggr\}\nonumber\\
+i\biggl\{ 4{\Phi_0}'\tilde{\phi}+\frac{2}{3}u\tilde{\nu}'+\frac{1}{3}u'\tilde{\nu}+\frac{2}{3}\frac{a'}{a}u\tilde{\nu} \biggr\}=0;\nonumber\\
\label{Eqs_pert}
\biggl\{\frac{1}{3}n^2\tilde{\nu} - a^2\Phi_0(m^2-\alpha\Phi_0^2)\tilde{\phi}-8\pi a^2\tilde{\delta\varepsilon}  \biggr\} +iu\{-\Phi_0'\tilde{\phi}+\frac{a'}{a}(\tilde{\nu}-\tilde{\lambda}) \}=0,
\end{eqnarray}

where we introduced a new independent function
\begin{eqnarray}
\nu=\lambda+\mu \Rightarrow & \mu=\nu-\lambda.
\end{eqnarray}
The terms in the curly brackets of the above equations refer to the zeroth and first orders of the WKB approximation, the second-order terms proportional to the squares of the first derivatives of the background functions or their second derivatives were discarded. At the same time, we retained in the zeroth-order terms, along with $n^2,u^2$, expressions like $a^2\times\{m^2,\tilde{\delta\varepsilon}\}$ due to the possible exponentially fast growth of the scale factor (for details, see \cite{WKB_22}).

Let us make two remarks regarding the obtained system \eqref{Eqs_pert}. First, the transition from the model with a scalar charged liquid to the model with a neutral liquid is carried out by a simple limit transition $e\to0$.
Second, the perturbations of the metric $\nu$ and $\lambda$ are dimensionless, as is the perturbation of the scalar potential $\phi$, and the perturbation of the energy density $\delta\varepsilon$ has the dimension of the energy density $\ell^{-2}$,
therefore, under scaling transformations of the model \eqref{Par_tilde} -- \eqref{tilde_t} these quantities are transformed in the same way as the corresponding unperturbed quantities. Further, since $a(\eta)$ is a dimensionless function invariant with respect to the scale transformation of the model, the time variable $\eta$ is transformed under scale transformations of the model in the same way as the cosmological time $t$ \eqref{tilde_t}, and, consequently, the wave number $n$ is transformed according to the law
\begin{equation}\label{tilde_n}
\tilde{n}=\frac{n}{k}.
\end{equation}
\subsection{WKB zero-approximation equations}
So, in the WKB zero-approximation we have a system of homogeneous algebraic equations with respect to $[\tilde{\lambda},\tilde{\nu},\tilde{\varepsilon},\tilde{\phi}]$:
\begin{eqnarray}\label{EQ-WHB_0}
\mbox{WKB}_0:\quad \left\{
\begin{array}{l}
\displaystyle u^2\tilde{\lambda}+\frac{1}{3}n^2\tilde{\nu}=0;\\
\displaystyle -\frac{1}{3}u^2\tilde{\nu}-a^2\Phi_0\bigl(m^2-\alpha\Phi_0^2-\frac{16\pi}{3}e^2\varepsilon_0\bigr)\tilde{\phi}+\frac{8\pi}{3} a^2(1-e^2\Phi_0^2)\tilde{\delta\varepsilon}=0;\\
\displaystyle \frac{1}{3}n^2\tilde{\nu} - a^2\Phi_0(m^2-\alpha\Phi_0^2)\tilde{\phi}-8\pi a^2\tilde{\delta\varepsilon}=0;\\
\displaystyle \bigl[n^2-u^2+a^2(m^2-3\alpha\Phi^2_0-8\pi e^2\varepsilon_0)\bigr] \tilde{\phi} -8\pi a^2e^2\Phi_0\tilde{\delta\varepsilon}=0.\\
\end{array}
\right.
\end{eqnarray}
Introducing the vector function $\mathbf{X}=[\tilde{\lambda},\tilde{\nu},\tilde{\delta\varepsilon},\tilde{\phi}]^T$, we write \eqref{ EQ-WHB_0} in the form of a matrix equation: 
\begin{equation}\label{Xx=0} \mathbf{A(u,n)\times X=0}, \end{equation} 
\begin{eqnarray}\label{A}
\mathbf{A(u,n)}=\quad \left(\begin{array}{cccc}
\displaystyle u^2 & \displaystyle \frac{1}{3}n^2 & 0 & 0\\[6pt]
\displaystyle 0 & \displaystyle-\frac{1}{3}u^2  & \displaystyle \frac{8\pi}{3} a^2(1-e^2\Phi_0^2) & \displaystyle -a^2\Phi_0\biggl(m^2-\alpha\Phi_0^2-\frac{16\pi}{3}e^2\varepsilon_0\biggr)\\[6pt]
\displaystyle 0  &\displaystyle  \frac{1}{3}n^2 & -8\pi a^2 & - a^2\Phi_0(m^2-\alpha\Phi_0^2)\\[6pt]
\displaystyle 0 & 0 & -8\pi a^2e^2\Phi_0 & n^2-u^2+a^2(m^2-3\alpha\Phi^2_0-8\pi e^2\varepsilon_0) \\[6pt]
\end{array}
\right).
\end{eqnarray}

A necessary and sufficient condition for a nontrivial system of linear homogeneous algebraic equations \eqref{Xx=0} is that the determinant of the matrix of this system $\mathbf{A(u)}$ \eqref{A} is zero:
\begin{equation}\label{detA=0}
\mathrm{Det}(\mathbf{A(u,n)})=0.
\end{equation}
Thus, we obtain a characteristic equation for the eikonal function, which in the theory of medium oscillations is usually called the \emph{dispersion equation}. This equation is an algebraic equation of the 3rd degree with respect to the square of the eikonal function. Therefore, we always have 3 pairs of symmetric solutions $u_\pm$, which describe oscillations propagating in opposite directions (advanced and retarded). Due to the specific form of the matrix $\mathbf{A}$, we have a doubly degenerate zero solution $u=0$, which, due to the first equation of the system \eqref{EQ-WHB_0}, leads to the result
\begin{equation}\label{nu=0}
u^2=0\Rightarrow \nu=0\Rightarrow \lambda+\mu=0,
\end{equation}
corresponding to admissible transformations of the metric.

To determine the remaining two pairs of symmetric solutions, we have an algebraic equation of the second degree with respect to $y\equiv u^2$:
\begin{eqnarray}\label{det=0}
\Delta(u^2,n^2)=\quad \left|\begin{array}{ccc}
\displaystyle-\frac{1}{3}u^2  & \displaystyle \frac{8\pi}{3} a^2(1-e^2\Phi_0^2) & \displaystyle -a^2\Phi_0\biggl(m^2-\alpha\Phi_0^2-\frac{16\pi}{3}e^2\varepsilon_0\biggr)\\[6pt]
\displaystyle  \frac{1}{3}n^2  & -8\pi a^2 & - a^2\Phi_0(m^2-\alpha\Phi_0^2)\\[6pt]
0 & -8\pi a^2e^2\Phi_0 & n^2-u^2+a^2(m^2-3\alpha\Phi^2_0-8\pi e^2\varepsilon_0) \\[6pt]
\end{array}
\right|=0\Rightarrow\\
\Delta(u^2,n^2)=\quad \left|\begin{array}{ccc}
\displaystyle -u^2  & \displaystyle \frac{1}{3}(1-e^2\Phi_0^2) & \displaystyle -a^2\Phi_0\biggl(m^2-\alpha\Phi_0^2-\frac{16\pi}{3}e^2\varepsilon_0\biggr)\\[6pt]
\displaystyle  n^2  & -1 & - a^2\Phi_0(m^2-\alpha\Phi_0^2)\\[6pt]
0  & -2e^2\Phi_0 & n^2-u^2+a^2(m^2-3\alpha\Phi^2_0-8\pi e^2\varepsilon_0)\\[6pt]
\end{array}
\right|=0. \nonumber
\end{eqnarray}

Let further $u^2_{1,2}$ -- roots of the dispersion equation \eqref{det=0}, then in the zeroth approximation we have 4 independent roots (not counting the trivial one):
\begin{eqnarray}
u^\pm_\alpha(n,\eta)=\pm \sqrt{u^2_\alpha}, & (\alpha=\overline{1,2}).
\end{eqnarray}
Substituting these solutions one by one into the equations \eqref{Xx=0}, we obtain the corresponding fundamental solutions of this system $\mathbf{X^\pm_\alpha(n,\eta)}$
\begin{equation}\label{X_ax=0}
\mathbf{A(u^\pm_\alpha(n,\eta),n)\times X^\pm_\alpha(n,\eta)=0},
\end{equation}
which are the solutions in the zero-order WKB approximation. To obtain the first-order solutions of the WKB approximation, it is necessary to use the found solutions in the system of linear ordinary differential equations of the first order, corresponding to the second curly brackets of the system \eqref{Eqs_pert}.

In this article, we will not calculate the roots of the dispersion equation \eqref{det=0} and find the zero-order WKB solutions \eqref{X_ax=0}, since this is an independent, rather complex computational problem \footnote{Note that the functions defining the matrix $\mathbf{A(u,n)}$ are solutions of nonlinear background equations.}, which we intend to study in a separate article. In this article, we will study some special cases and discuss the general properties of the WKB solution.

\subsection{The case of a neutral ideal fluid $\mathbf{S^{(0)}}$}
Assuming for a neutral fluid $e^2\equiv0$, we obtain for the dispersion equation \eqref{det=0}
\begin{eqnarray}\label{det=0,e=0}
\Delta(u^2,n^2)=\quad \left|\begin{array}{ccc}
\displaystyle -u^2 & \displaystyle \frac{1}{3} & \displaystyle -a^2\Phi_0(m^2-\alpha\Phi_0^2)\\[6pt]
\displaystyle n^2 & -1 & - a^2\Phi_0(m^2-\alpha\Phi_0^2)\\[6pt]
0 & 0 & n^2-u^2+a^2(m^2-3\alpha\Phi^2_0)\\[6pt] \end{array} \right|=0, \end{eqnarray} 
from where we find 
\begin{eqnarray}\label{u=1/3n} u^2-\frac{1}{3}n^2=0\Rightarrow & \displaystyle u^\pm=\pm\frac{n }{\sqrt{3}};\\ 
\label{us=sqrt} n^2-u^2+a^2(m^2-3\alpha\Phi_0^2)=0\Rightarrow & \displaystyle u_s^\pm=\pm\sqrt{n^2+a^2(m^2-3\alpha\Phi_0^2)} 
\end{eqnarray} 

The solutions $u^\pm$ \eqref{u=1/3n} correspond to a pair of undamped waves propagating at the speed of sound $v_s=1/3$ in opposite directions, and in this mode $\tilde{\phi}=0$, $\tilde{\lambda}=\tilde{\delta\varepsilon}/n^2$. Thus, these are ordinary sound vibrations of a neutral liquid. The solutions $u_s^\pm$ for $m^2\geqslant 3\alpha\Phi^2_0$ also correspond to a pair of undamped waves,
propagating with the phase velocity
\[v_f=\sqrt{1+\frac{a^2}{n^2}(m^2-3\alpha\Phi^2_0).}\]
For $m^2< 3\alpha\Phi^2_0$, a sector of disturbance wave numbers arises, in which the eikonal function takes imaginary values:
\begin{equation}\label{u=Im}
n^2-a^2(3\alpha\Phi^2_0-m^2)<0\Rightarrow u^\pm_s=\pm i\sqrt{a^2(3\alpha\Phi^2_0-m^2)-n^2}\equiv \pm i\gamma(n,\eta),
\end{equation}
where $\gamma(n,\eta)$ is the increment/decrement of the increase/decrease of oscillations.  Non-zero perturbations $\lambda$, $\delta\varepsilon$ and $\phi$ represent pairs of standing decaying and growing modes:
\begin{equation}
\displaystyle \lambda\sim \displaystyle C_{-}\mathrm{e}^{\displaystyle inz-\int\gamma(n,\eta)d\eta}+C_{+}\mathrm{e}^{\displaystyle inz+\int\gamma(n,\eta)d\eta},
\end{equation}
i.e., at time $\eta_0$, if it exists, --
\begin{equation}\label{eta_0}
a^2(\eta_0)(3\alpha\Phi^2_0(\eta_0)-m^2)=n^2.
\end{equation}
in one of the modes \eqref{eta_0} becomes unstable. Note also that in the studied model of neutral fluid the eikonal functions $u^\pm_s$ according to \eqref{u=Im} can be either real or purely imaginary. Therefore, on different time intervals the disturbances can be either free waves with local frequency $\omega_t(t)$ or standing growing and damped oscillations with growth increment $\gamma_t(t)$\footnote{This and more complex types of behavior in the case of a scalar charged fluid are indicated in \cite{Yu_GC_3_22}.}.
For the correct interpretation of the above results in the expression \eqref{u=Im} it is necessary to pass from the time variable $\eta$ to the cosmological time $t$ using the formulas \eqref{a(eta)} and introducing the \emph{integral increment of disturbance growth} $\chi(t)$ and the oscillation phase $\Psi(t)$
\begin{eqnarray}\label{chi(t)=}
\chi(t) \equiv \Re\int\gamma d\eta= \int\limits_{t_0}^t\gamma_t(t)dt, & \gamma_t(t)=\mathrm{e}^{-\xi(t)}\Re(\gamma(t); \\
\label{Psi(t)=}
\Psi(t)\equiv \Im\int\gamma d\eta=\int\limits_{t_0}^t\omega_t(t)dt , & \omega_t(t)=\mathrm{e}^{-\xi(t)}\Im(\gamma(t)).
\end{eqnarray}
so that the amplitude of the perturbation of the scalar field changes according to the law
\begin{equation}\label{phi(t)}
\tilde{\phi}(t)\backsimeq \mathrm{e}^{\chi(t)+i\Psi(t)}; \; \phi(z,t)\backsimeq \mathrm{e}^{\chi(t)}\cdot \mathrm{e}^{i(kz+\Psi(t))}
\end{equation}
Note that according to the definition \eqref{chi(t)=} the growth of the integral increment of disturbances stops at the moment of time when $\Re(\gamma(t))=0$.

Fig. \ref{fig16} -- \ref{fig17} shows the evolution of the geometric factors $\xi(t)$ and $H(t)$ in an initially unstable model with an ultrarelativistic neutral fluid with parameters $\mathbf{P^{(0)}_0}$ \eqref{P_00} and initial conditions $\mathbf{I_1}$ \eqref{I_1} in the case of wave number $n=1$, as well as the evolution of the local increment $\gamma_t(t)$ and the local frequency $\omega_t(t)$ of oscillations for the same model.

Fig. \ref{fig18} -- \ref{fig19} shows the evolution of the integral increment of perturbation growth in an initially unstable model with an ultrarelativistic fluid with parameters $\mathbf{P^{(0)}_0}$ \eqref{P_00} and initial conditions $\mathbf{I_1}$ \eqref{I_1} in the case of wave number $n=1$. In this case, Fig. \ref{fig18} shows the graph $\xi(t)$ against the background of the graph $H(t)$, and Fig. \ref{fig19} shows the same graph against the background of the graph of the function $3\alpha\Phi^2(t)-m^2$,
which determines the increment of oscillation growth \eqref{u=Im}. From these graphs it is evident that the end of the growth of disturbances is connected, firstly, with the process of transition of the model from an unstable state to a stable one, and, secondly, with the change of sign of the function $3\alpha\Phi^2(t)-m^2$. After the growth of the integral increment of oscillations ceases, the eikonal functions \eqref{u=Im} become real, which corresponds to free oscillations of the scalar field in the form of pairs of retarded and advanced potentials with a fixed amplitude.

\TwoFigsReg{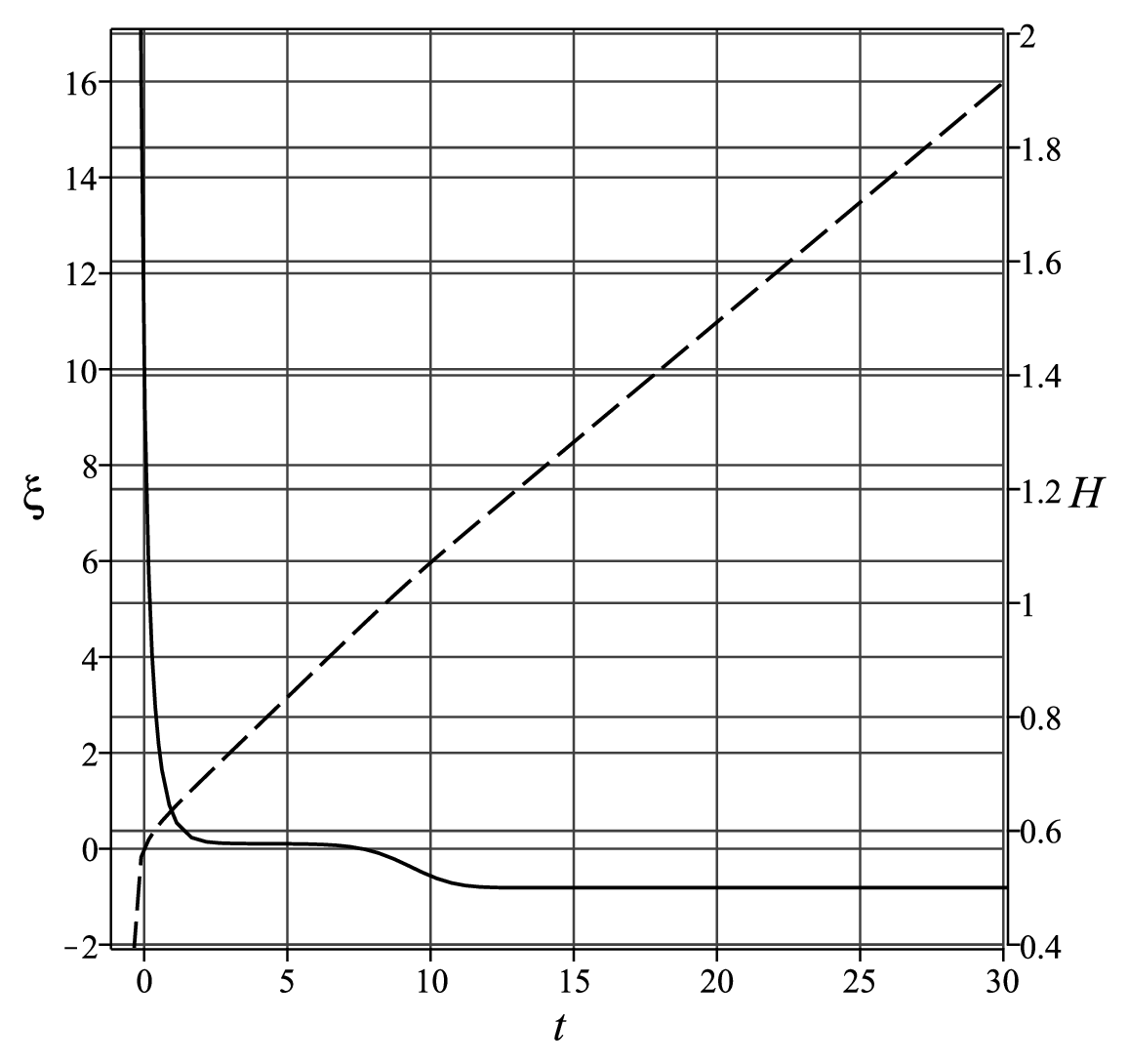}{6}{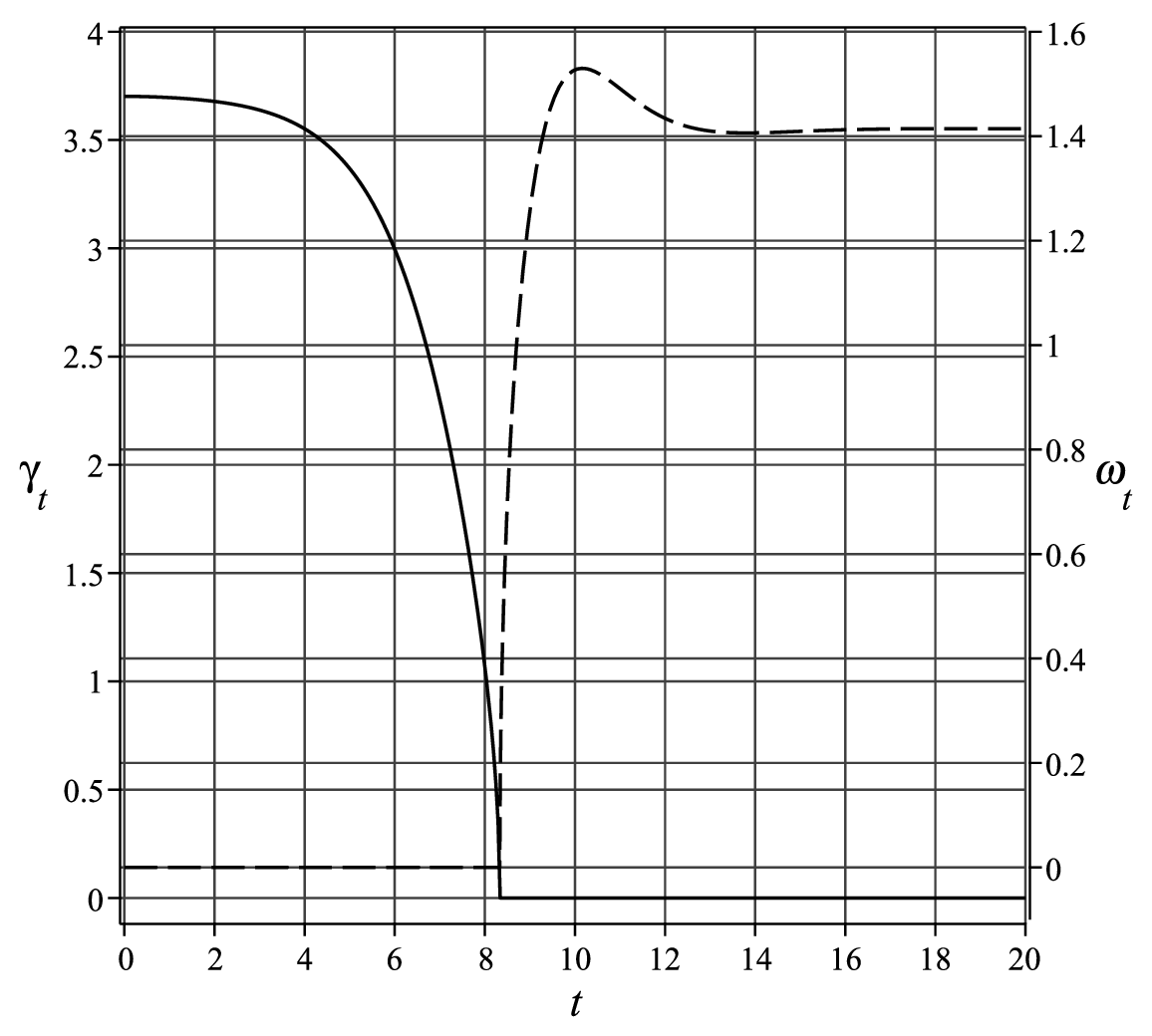}{6}{\label{fig16}Evolution of geometric factors $\xi(t)$ (dashed line) and the Hubble parameter $H(t)$ (solid line ) in the $\mathbf{S^{(0)}}$ model: $\mathbf{P}=\mathbf{P_0}$; $\mathbf{I}=\mathbf{I_1}$.}{\label{fig17}Evolution of the increment $\gamma_t(t)$ (solid line) and oscillation frequency $\omega_t(t)$ (dashed line) in the model $\mathbf{^{(0)}}$: $\mathbf{P}=\mathbf{P_0}$; $\mathbf{I}=\mathbf{I_1}$; $n=1$.\\} 
\TwoFigsReg{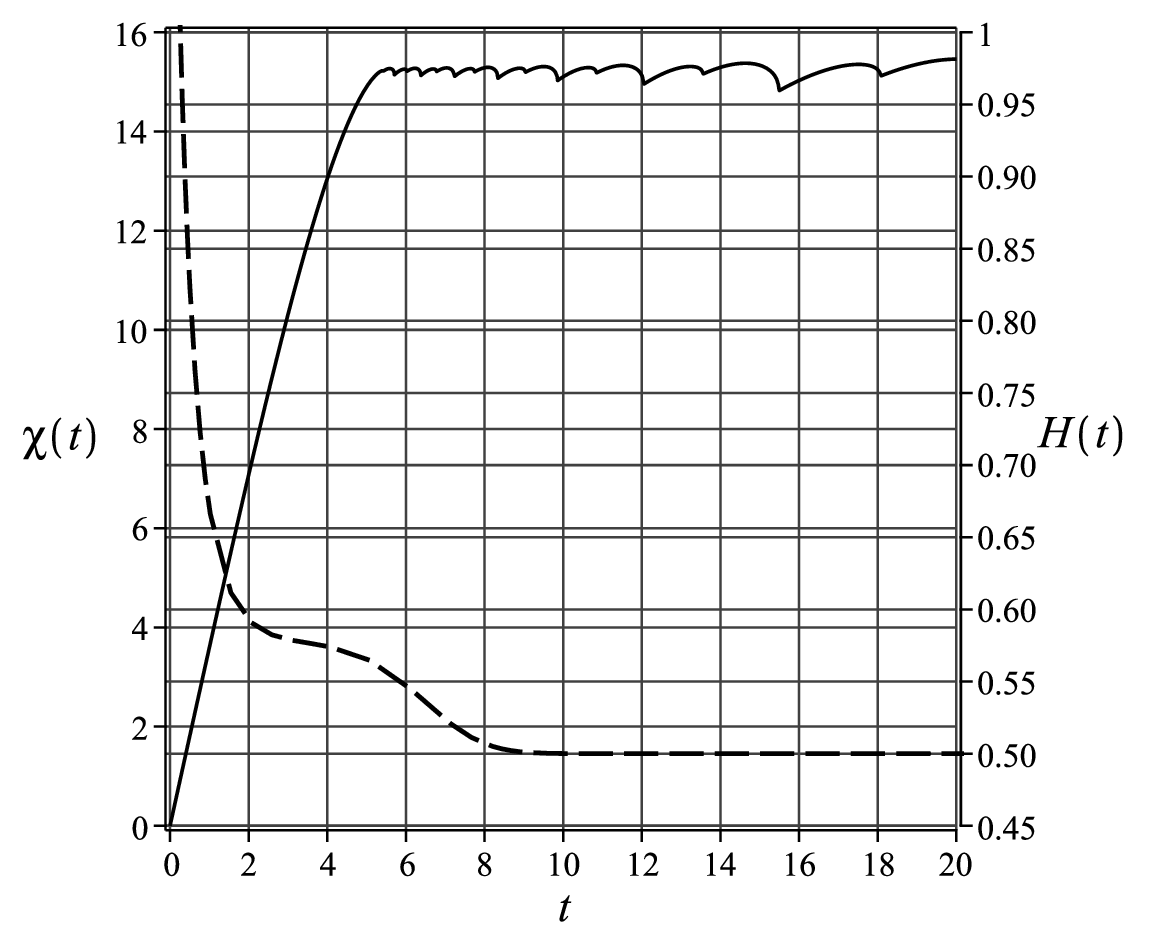}{6}{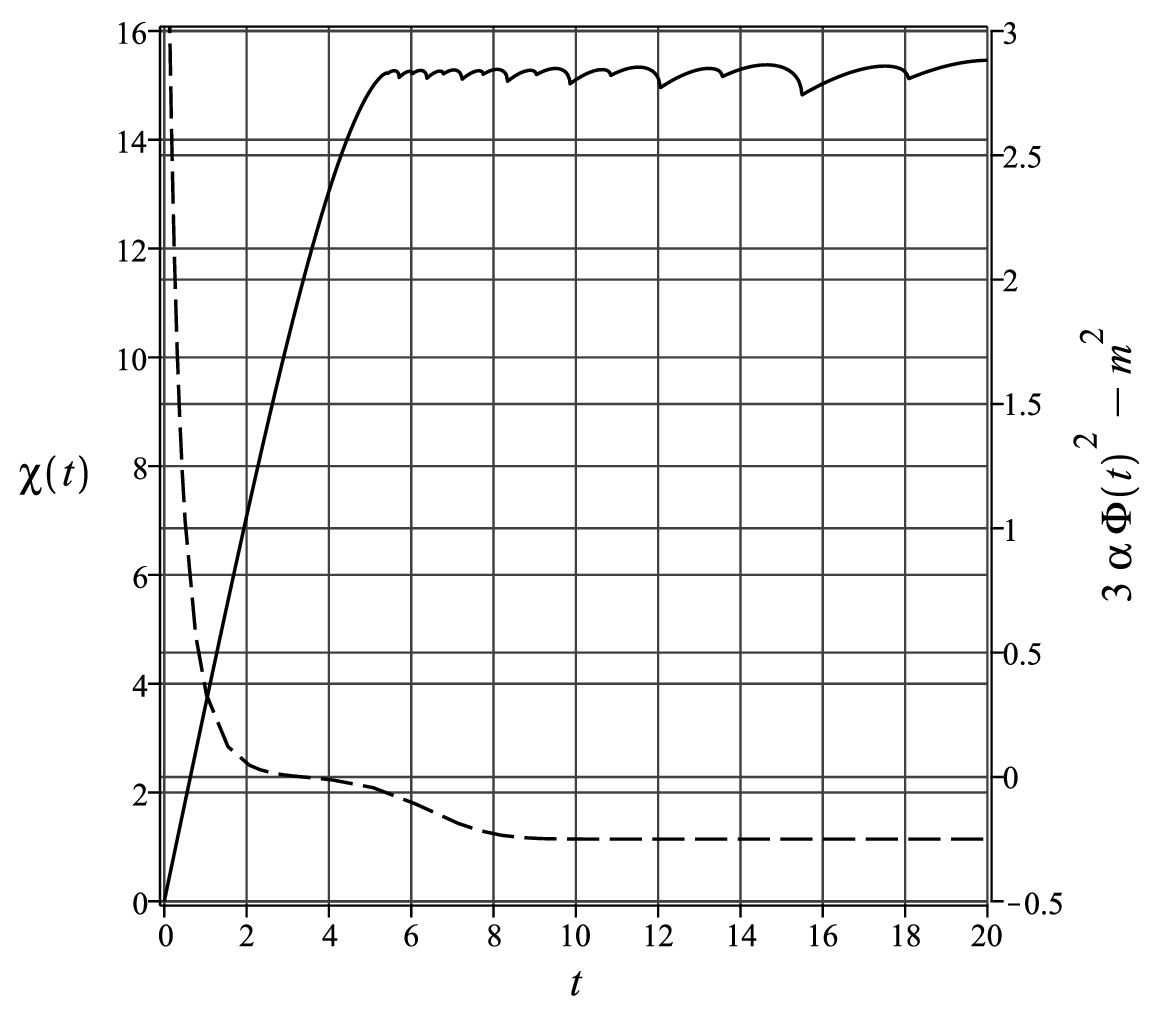}{6}{\label{fig18}Evolution of the integral increment $\chi(t)$ (solid line) and Hubble parameter $H(t)$ (dashed line) in the model $\mathbf{M^{(0)}_0}$: $\mathbf{P}=\mathbf{P^{(0)}_0}$; $\mathbf{I}=\mathbf{I_2}$; $n=1$.}{\label{fig19}Evolution of the integral increment $\chi(t)$ (solid line) and function $3\alpha\Phi^2(t)-m^2$ (dashed line) in the model $\mathbf{M_0}$: $\mathbf{P}=\mathbf{P_0}$; $\mathbf{I}=\mathbf{I_2}$; $n=1$.\\} %

\subsection{Expansion of the dispersion relation in a series in terms of the smallness of the scalar charge $e$}
We now consider the dispersion relation \eqref{det=0} for small values ??of the scalar charge parameter, restricting ourselves to terms of the second order in terms of $e$. In the zeroth and first approximations of smallness of $e^2$ we find from
\eqref{det=0}:
\begin{eqnarray}\label{Det_0}
\!\!\!\Delta_0\equiv \left.\Delta(u^2,n^2)\right|_{e=0}=\left(u^2-\frac{n^2}{3}\right)(-u^2+n^2+a^2\Phi_0(m^2-3\alpha\Phi_0^2));\\
\!\!\!\Delta_1=\frac{\Phi_0e^2}{3}\biggl[6\Phi_0 a^2(u^2+n^2)(m^2-\alpha\Phi_0^2) -3\Phi_0^2u^2(3\alpha\Phi_0^2-m^2)-\Phi_0 n^2(u^2-n^2)-24\pi\varepsilon_0 a^2\biggl(u^2-\frac{n^2}{3}\biggr), \end{eqnarray} 
so 
\begin{equation} \Delta(u^2,n^ 2)\approx \Delta_0(u^2)+\Delta_1(u^2).
\end{equation} 
Further assuming 
\[\displaystyle u^2\equiv x; \qquad \displaystyle x=x_0+x_1 \Rightarrow u^2=u^2_0+\delta u^2,\]
expand the solution of the dispersion equation \eqref{det=0} into a Taylor series, assuming
\begin{eqnarray}\label{Det_0=0}
\Delta_0(x_0)=0; \\
\label{Det_1=0}
\Delta(x_0+x_1)\approx \Delta_0(x_0+x_1)+\Delta_1(x_0)=0.
\end{eqnarray}
Solving the zeroth order dispersion relation \eqref{Det_0=0}, we find the modes \eqref{u=1/3n} and \eqref{us=sqrt}, the first of which describes the sound vibrations of an ultrarelativistic liquid. The unstable mode, as we have seen, is the second \eqref{us=sqrt}, using which in \eqref{Det_1=0}, we find:
\begin{eqnarray}\label{Det0_1=}
\Delta_0(u^2_0+\delta u^2)= -\frac{1}{3}[2n^2-3a^2\Phi_0(3\alpha\phi_0^2-m^2)] \cdot \delta u^2;  \nonumber\\
\label{Det_1_0=}
\Delta_1(u^2_0)=\frac{2e^2\Phi_0^2 a^2}{3}\biggl\{6\Phi_0n^2(m^2-\alpha\Phi_0^2)-4\pi\varepsilon_0[2n^2-3a^2\Phi_0(3\alpha\Phi_0^2-m^2)] \nonumber\\
+3\Phi_0^2 a^2(3\alpha^2\Phi_0^4-4\alpha m^2\Phi_0^2+m^4)\biggr\},  \nonumber\\
u^2\approx n^2-a^2\Phi_0(3\alpha\Phi_0^2-m^2)+\frac{3\Delta_1(u^2_0)}{2n^2-3a^2(3\alpha\Phi_0^2-m^2)}
\end{eqnarray}

Fig. \ref{fig20} -- \ref{fig23} shows the dependence of the square of the eikonal function, as well as the evolution of this square of the eikonal and the contributions $u^2(t)$ and the contributions to it $u^2_0(t)$ and $\delta u^2(t)$ according to the above formulas. 
\TwoFigsReg{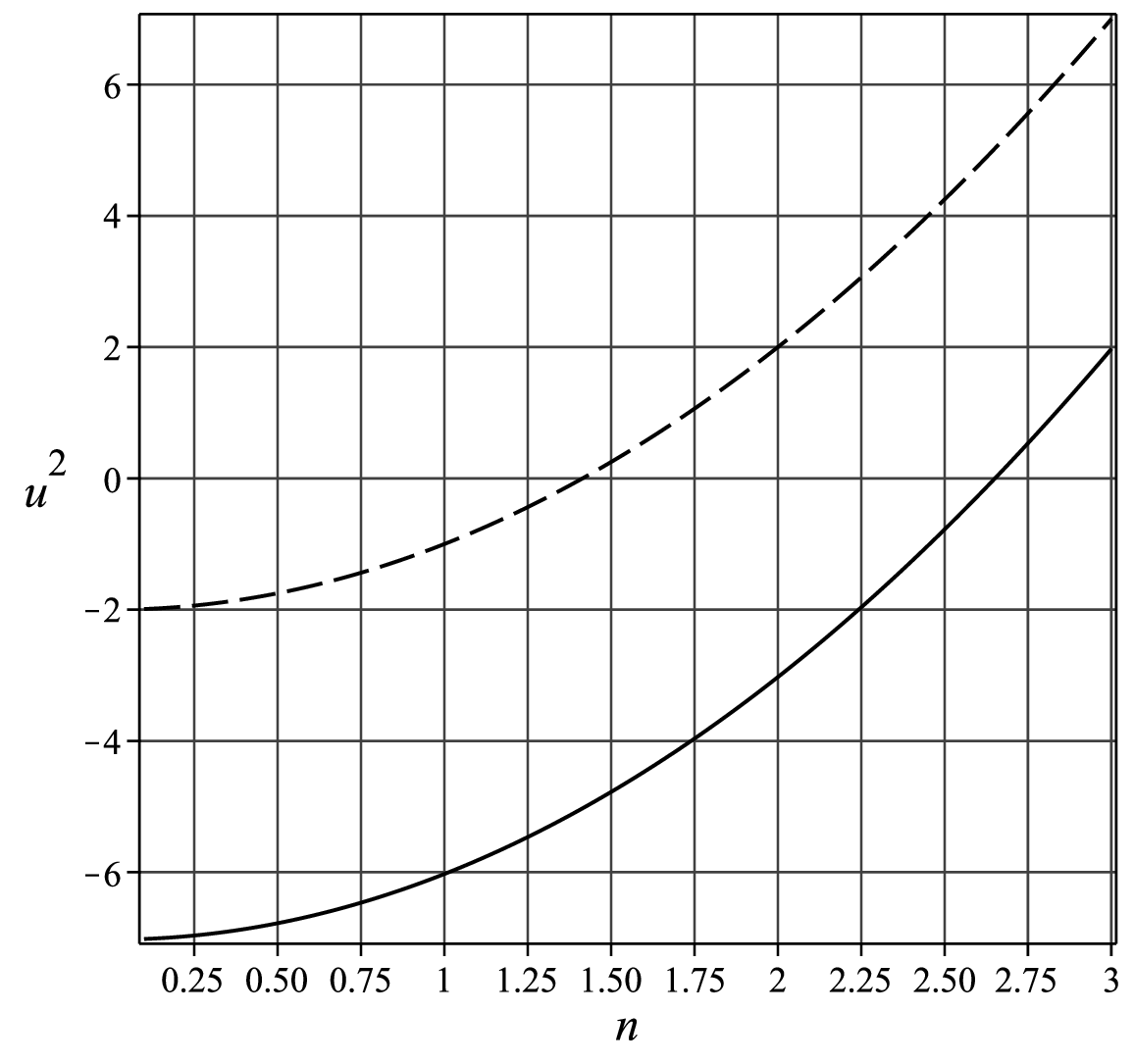}{6}{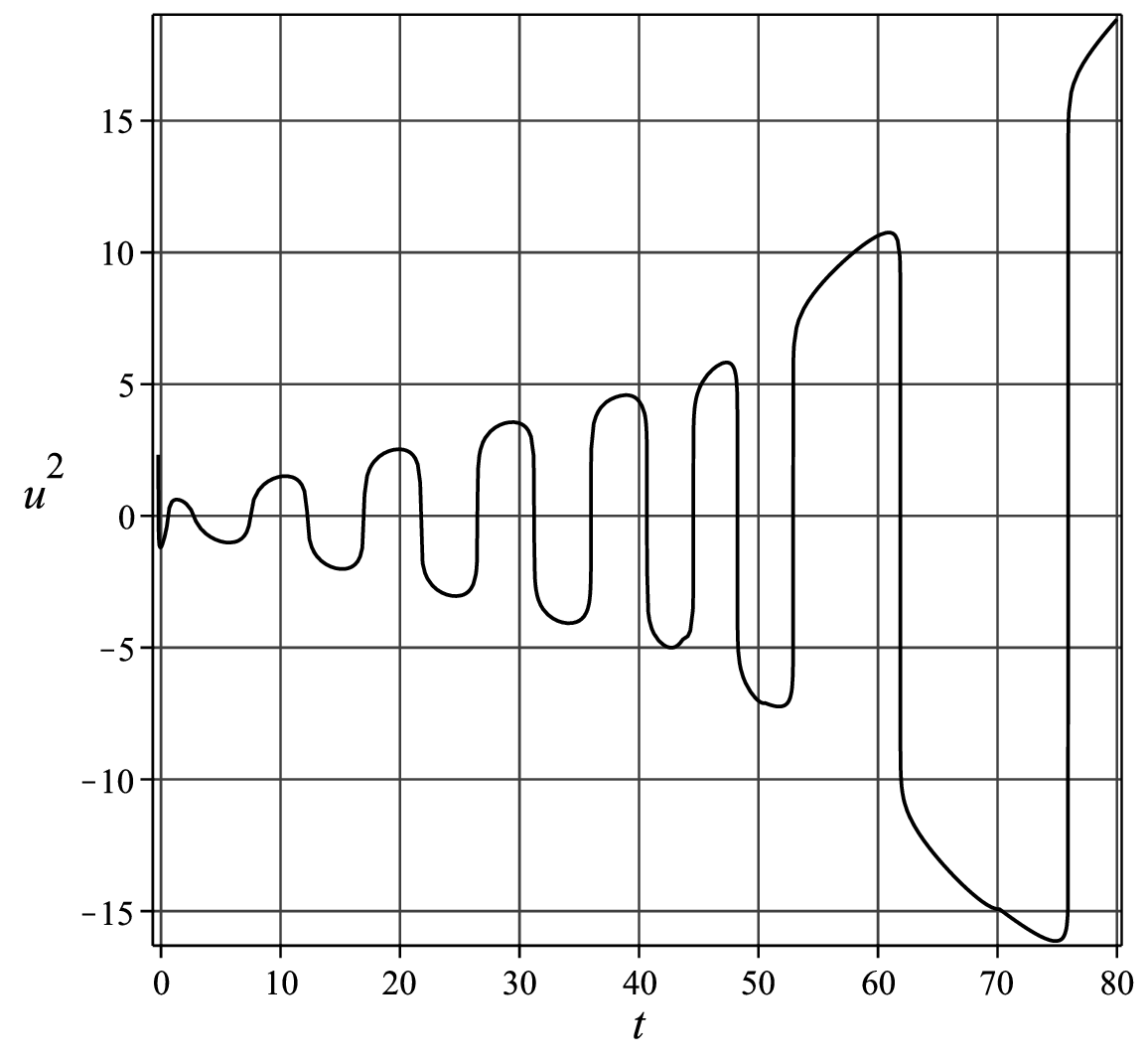}{6}{\label{fig20}Dependence of eikonal squares $u^2(n)$ \eqref{Det_1_0=} and \eqref{us=sqrt} (dashed line) 
in the model  $\mathbf{S}$: $\mathbf{P}=[1,1,1,1]$  for $a=\Phi_0=\varepsilon_0=1$}{\label{fig21}Evolution of the squared eikonal $u^2$ \eqref{Det_1_0=} 
in the model $\mathbf{S}$: $\mathbf{P}=[1,1,1,1]$, $\mathbf{I}=[0.99,0,0.5,1]$.\\}
\TwoFigsReg{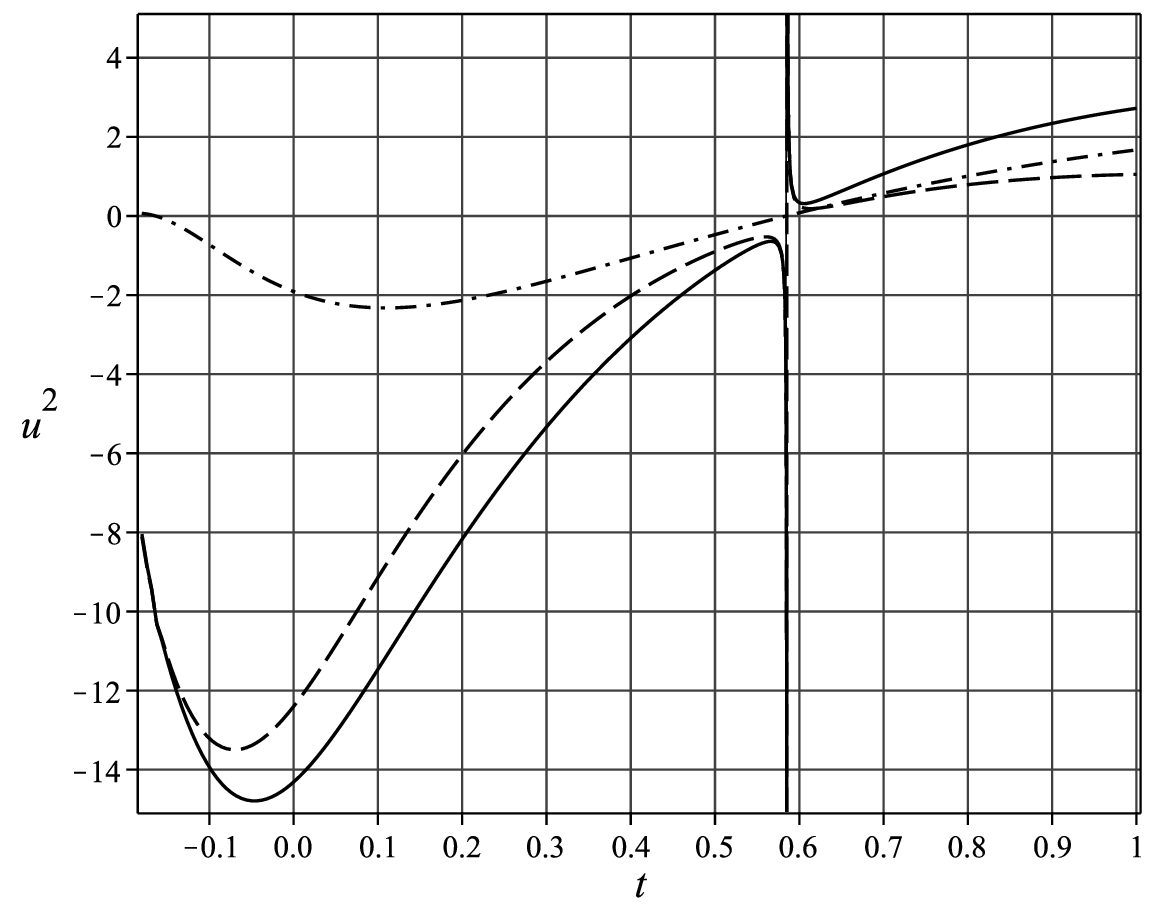}{6}{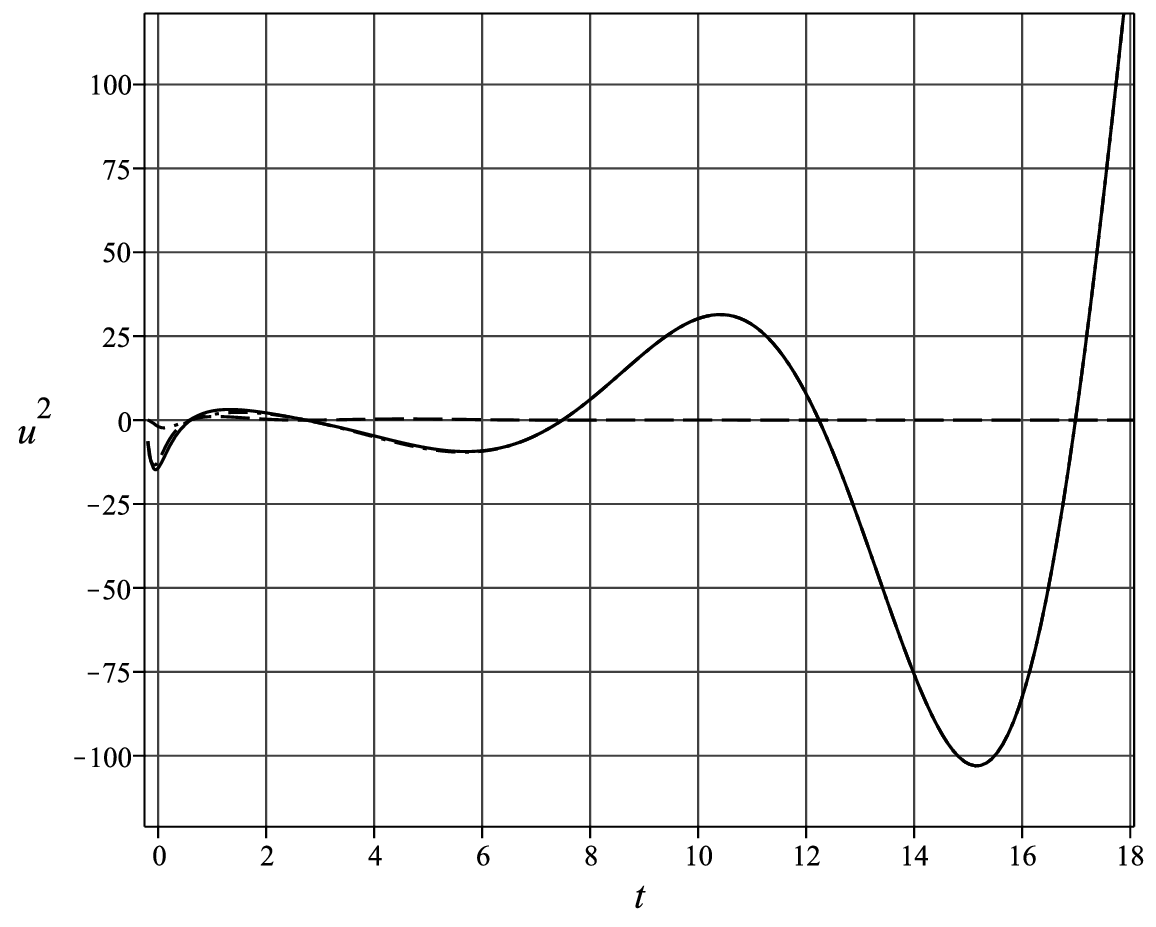}{6}{\label{fig22}Evolution (on a small scale) of the eikonal squares $u^2$ \eqref{Det_1_0=} (solid line), $\delta u^2$ (dashed line) and $u_0^2$ \eqref{us=sqrt} (dash-dotted line) in the $\mathbf{S}$ model: $\mathbf{P}=[1,1,1,1]$, $\mathbf{I}=[0.99,0,0.5,1]$.}{\label{fig23}Evolution (on a large scale) of the eikonal squares $u^2$ \eqref{Det_1_0=} (solid line), $\delta u^2$ (dashed line) and $u_0^2$ \eqref{us=sqrt} (dash-dotted) in the $\mathbf{S}$ model: $\mathbf{P}=[1,1,1,1]$, $\mathbf{I}=[0.99,0,0 .5,1]$.\\}

From the graphs presented above in Fig. \ref{fig20} -- \ref{fig23} one can notice two important features of the instability of oscillations, i.e., the negativity of the square of the eikonal $u^2<0$. First, the scalar charge factor of the liquid becomes dominant at the earliest stages of evolution, while at later stages the instability of the neutral liquid dominates. Second, the function $u^2(t)$ at later stages has the form of anharmonic oscillations with increasing amplitude and increasing period. Note that at $u^2>0$ the oscillations of the system have the form of pairs of traveling waves with constant amplitude, while at $u^2<0$ the oscillations turn into pairs of standing growing and damping harmonics. The amplitude of the standing waves after the termination of the growth phase is inherited each time by the traveling waves. This process repeats itself, transforming the evolution of the Universe into an alternating chain of increasingly longer stages of disturbance growth and stages of traveling waves with ever greater amplitude. Eventually, one of the stages of disturbance growth must reach a nonlinear stage, at which this quasi-periodic process can end, precisely at the stage of standing oscillations, thereby leading to the formation of the structure of the Universe.

\subsection{Is instability of the vacuum-field model $\mathbf{S^{(00)}}$ possible with respect to longitudinal perturbations?}
In the work of M. Yu. Khlopov, B. A. Malomed and Ya. B. Zeldovich \cite{Khlopov}, to which M. Yu. Khlopov kindly pointed out to the Author, within the framework of the Newtonian model, a result was obtained on the instability of the \emph{spatially homogeneous static background} with respect to perturbations of the classical scalar field and on the occurrence of standing growing oscillation modes. These results are qualitatively similar to those given above, although they are based on a semiclassical model of gravity with a classical Newtonian potential. In this case, \cite{Khlopov} studies a system consisting of one linearized Einstein equation $^4_4$ with the energy-momentum tensor of a scalar field with a self-action constant and a scalar field equation; there is no liquid or dust in the model. The model also does not take into account the Einstein equations $\{4,\alpha\}$, which are precisely what lead to the above-mentioned incorrectness of the vacuum-field model.

Let us consider the question of the possibility of instability of the vacuum-field cosmological model $\mathbf{S^{(00)}}$, i.e., in the absence of an ideal fluid in it. As we noted in the \ref{pert_vac} section, a correct consideration of perturbations in the vacuum-field model is possible only in the case when the system is at one of the singular points: $\Phi_0(m^2-\alpha\Phi_0^2)=0$ \eqref{Phi_0Phi_0=0}. The transition to this case is carried out, firstly, when the conditions $\varepsilon_0\equiv0$, $e\equiv0$ are satisfied and by eliminating from the determinant \eqref{det=0} the second column corresponding to the perturbation of the energy density of the ideal fluid $\tilde{\delta\varepsilon}$, and the corresponding second row. In this case, we obtain a second-order determinant and the corresponding dispersion equation of the zeroth WKB approximation:
\[u^2[n^2-u^2+a^2(m^2-3\alpha\Phi_0^2)]=0.\]
Thus, we again obtain one zero mode of oscillations and a dispersion equation of the standard form for the theory of oscillations:
\[u^2=n^2+a^2(m^2-3\alpha\Phi_0^2).\]
It follows that for
\[m^2-3\alpha\Phi_0^2\geqslant0\Rightarrow u^2>0\]
we obtain two free waves propagating in opposite directions. Instability of the cosmological model can arise only in the case $u^2<0$, which is possible only for $m^2<3\alpha\Phi_0^2$.
As follows from the results of the qualitative theory of dynamic systems \eqref{attract_eq}, when the additional condition $\Lambda^2-4m^2<0$ is satisfied, the cosmological model has only one stable singular point corresponding to the zero scalar potential $\Phi_0=0$. But in this case, the dispersion equation gives $u^2=n^2+a^2m^2$, i.e., it describes a pair of undamped waves. Instability of perturbations can arise only in the case of an unstable state of the unperturbed cosmological model, which we discussed in the \ref{pert_vac} section. The graphs in Fig. \ref{fig16} -- \ref{fig17} demonstrate the fact that the growth of perturbation amplitudes occurs precisely at the unstable stage of evolution.

Since the dispersion relation in the vacuum-field model $\mathbf{S^{(00)}_0}$ formally coincides with the dispersion relation in the WKB approximation \eqref{us=sqrt} for the mathematically correct model $\mathbf{S^{(0)}_0}$ with a neutral fluid, we can use this model to study longitudinal disturbances by substituting the background functions $\Phi(t)$ and $a(t)$ for the model $\mathbf{S^{(0)}_0}$ into the dispersion relations. Further, according to \eqref{u=Im} and \eqref{chi(t)=} the increment/decrement of the growth/damping of disturbances in terms of cosmological time $t$ is equal to
\begin{eqnarray}\label{gamma_t}
\gamma_t=\Re\sqrt{3\alpha\Phi_0^2(t)-m^2-\frac{n^2}{a^2(t)}} \equiv \Re\sqrt{3\alpha\Phi_0^2(t)-m^2-n^2\exp(-2\xi(t))},\\
\label{omega_t}
\omega_t=\Im\sqrt{3\alpha\Phi_0^2(t)-m^2-\frac{n^2}{a^2(t)}} \equiv \Im\sqrt{3\alpha\Phi_0^2(t)-m^2-n^2\exp(-2\xi(t))}.
\end{eqnarray}
Using a simple background model with two inflationary stages of expansion \eqref{a(eta)} and an instantaneous transition from the state of unstable inflation to the stable phase, the condition for the emergence of instability of the mode with wave number $n$ according to \eqref{gamma_t} can be written as:
\begin{equation}\label{unst_cond}
\gamma^2_t\geqslant 0\Rightarrow t_1> t\geqslant t_g(n) \equiv \sqrt{\frac{3}{\Lambda_0}}\ln\frac{n}{\sqrt{2}m},
\end{equation}
where $t_1$ is the moment of the end of the first inflationary (unstable) expansion phase with the Hubble parameter $H_1=\sqrt{\Lambda_0/3}$ and the transition to the second inflationary (stable) expansion phase with the Hubble parameter $H_2=\sqrt{\Lambda/3}$ ($H_2<H_1$). From \eqref{unst_cond} follows the necessary condition for the existence of the instability phase of the perturbation mode with the wave number $n$:
\begin{equation}\label{ness_cond}
t_1>t_g(n).
\end{equation}
In this case, the perturbation mode with wave number $n$ grows over time $\delta t=t_1-t_g(n)$ and grows during this time by
\begin{equation}\label{max_phi}
\frac{\phi_{max}}{\phi_0}\backsimeq \exp\biggl(\ \int\limits_{t_g}^{t_1} \gamma_tdt\ \biggr).
\end{equation}

Thus, the following statement is true.
\begin{stat}\label{u^2>0}\hskip -4pt{ .}
A vacuum-field cosmological model in a stable inflationary state is stable with respect to longitudinal scalar perturbations. Unstable modes can exist only in an unstable state of an unperturbed cosmological system.
\end{stat}

Note that this quite natural result is implicitly contained in the work \cite{TMF_20}. Summarizing this section, we note that although the model of the article \cite{Khlopov} is mathematically incorrect, it correctly predicts the main features of the scalar-gravitational instability of the Universe. For this model to be correct, it is necessary to add, firstly, at least a little ideal fluid, and, secondly, to make the background state non-stationary, subjecting it to Einstein's equations. Obviously, without taking this factor into account, the authors of the article \cite{Khlopov} could not have foreseen the quasi-periodic nature of the development of scalar-gravitational instability and associated the times of the beginning and end of the development of instability with the instability of the inflationary expansion itself and with the time of the transition of the inflationary expansion from an unstable regime to a stable one. Note also that the property of the eikonal function to be either purely real or purely imaginary is determined by the used model of matter, leading to a degenerate second-order dispersion equation. In a rigorous model of scalar charged matter, the order of the dispersion equation increases and its roots correspond to more complex oscillation modes \cite{WKB_22}.

Fig. \ref{fig24} shows the dependencies
\[\Delta t\equiv t_1(\Lambda_0)-t_g(\Lambda_0,n)\]
for the parameters of the model $\mathbf{S^{0}}$ for $\alpha=m=1,k=1/3$. The existence of an unstable phase is determined by the condition $\Delta t>0$, and its duration is equal to $\Delta t$.

\Fig{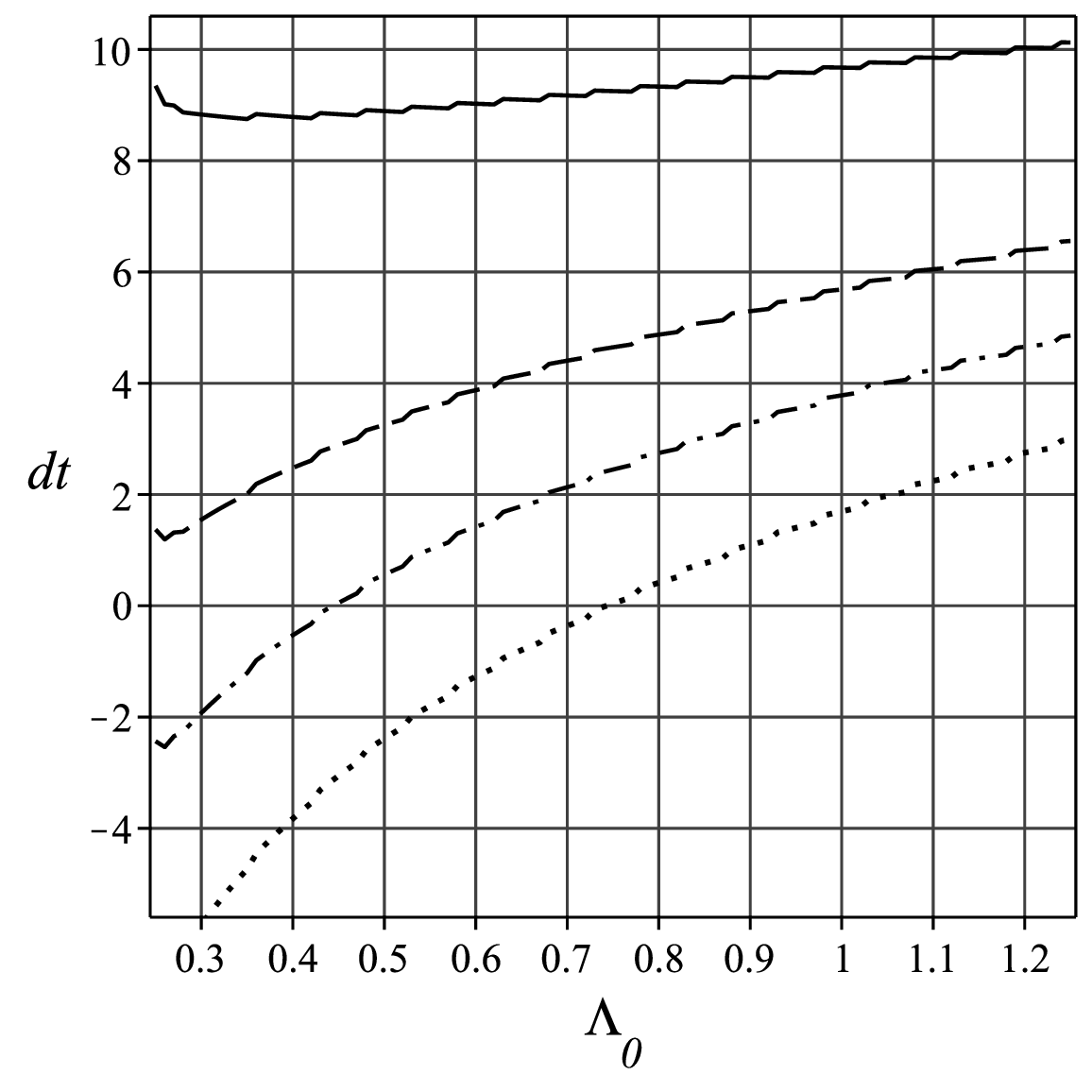}{8}{8}{\label{fig24} Graphs of $\Delta t(\Lambda_0,n)$ \eqref{unst_cond}: $n=1$ -- solid line, $n=10$ -- dashed line, $n=30$ -- dash-dotted line, and $n=100$ -- dotted line for the $\mathbf{S^{0}}$ model with $\alpha=m=1,k=1/3$.}
When scaling to the parameters of the $SU(5)$ field-theoretical model \eqref{SU5} with the similarity coefficient $\zeta=10^5$, we obtain $\Delta t \sim 10^6t_{pl}$ when $\Lambda_0$ varies within $10^{-11}\div10^{-10}$ in Planck scale.

\section{Conclusion}
In conclusion of the article we list its main results.

\begin{itemize}
\item A phenomenological model of an ideal fluid with a scalar charge is formulated. At the same time, macroscopic scalars of pressure $p$ and charge density $\sigma$ with properties similar to the properties of the
corresponding quantities in the microscopic theory are constructed.

\item A closed normal system of autonomous differential equations is formulated that describe the evolution of a spatially flat Friedmann Universe filled with an ideal scalar-charged fluid with a Higgs field.
The Cauchy problem is formulated and the transformation properties of the model are investigated. Based on this basic model, a model with a neutral fluid and a vacuum-field model are constructed and the rules for the transition between them are found.
\item A qualitative analysis of the obtained dynamic systems is carried out and numerical cosmological models based on these systems are constructed. It is shown that models with charged and neutral liquids give similar results
outside critical points, but differ significantly at critical points.
\item A mathematical model of plane longitudinal scalar-gravitational perturbations of the Friedmann ideal charged liquid with Higgs interaction is formulated. It is shown that in the absence of liquid, i.e., in
the vacuum-field model, gravitational perturbations do not arise. Perturbations of the scalar field are possible only in those cases when in the unperturbed state the cosmological system is at singular points.
For these cases, exact solutions of the field equation are found, expressed in Bessel functions of the first and second kind and describing damped oscillations in the case of a stable unperturbed state and growing oscillations in the case of an unstable unperturbed state.
\item A WKB theory of plane scalar-gravitational perturbations is constructed: dispersion equations are obtained in general form and solved in the case of a neutral fluid. In this case, expressions are obtained through the basic background
functions for the local frequency and growth increment of oscillations, as well as the integral increment, and graphs of the evolution of these quantities are presented. It is shown that only free wave or growing standing oscillation modes are possible during the evolution.
\item By expanding the dispersion equation in terms of the smallness of the scalar charge, approximate expressions are obtained for the eikonal function, defined through the basic background functions. Using numerical modeling, the evolution
of the eikonal function is investigated and it is shown that the square of this function evolves like a quasiperiodic function with increasing amplitude and period. The indicated periods represent a chain of stages of free waves and growing standing oscillations.
\item Perturbations in the WKB approximation in a neutral fluid are studied. It is shown that local formulas of perturbation evolution correspond to the model of the article \cite{Khlopov}. However, the times of the beginning and end of the instability phase are determined and it is shown that instability can develop only at the unstable inflationary stage of the expansion of the Universe.
\end{itemize}

Thus, the goals of the study formulated in the Introduction have been achieved. In the next article, we plan to study spherical perturbations based on the formulated model, trying to simplify the excessively cumbersome mathematical model of the formation of supermassive Black Holes, studied in \cite{TMF_23_1}.
\section*{Acknowledgments}
The author is grateful to the participants of the seminar of the Department of Relativity and Gravitation Theory of Kazan University for useful discussion of the work. The author is especially grateful to professors S.V. Sushkov and A.B. Balakin for very useful comments and discussion of the features of modern modifications of the theory of gravity as applied to cosmology and astrophysics. The author is also grateful to the participants of the VI International Winter School-Seminar "Petrovskie Readings-2023" for a fruitful discussion of the report (11/29/23), which influenced the topic of the article, especially to Academician \fbox{A.A. Starobinsky} and Professor K.A. Bronnikov.

\section*{Founding}
The work was carried out at the expense of a subsidy allocated as part of the state support of Kazan (Volga Region) Federal University in order to increase its competitiveness among the world's leading scientific and educational centers.

\end{document}